%% file: PiKPvsMulti13TeV.tex
\documentclass[ALICE,manyauthors]{cernphprep}

\usepackage[comma,square,numbers,sort&compress]{natbib}
\usepackage{comment}
\usepackage{hyperref}
\usepackage{lineno}
\usepackage{xspace}
\usepackage{xcolor}
\usepackage{graphicx}
\usepackage{dcolumn}
\usepackage{bm}
\usepackage{color}
\usepackage{hyperref}
\usepackage{multirow}
\usepackage{multicol}
\usepackage{booktabs}
\usepackage{epstopdf}
\usepackage{rotating}
\usepackage{tabularx}
\usepackage{epstopdf}
\usepackage{afterpage}
\usepackage[T1]{fontenc}
\newcolumntype{Y}{>{\centering\arraybackslash}X}
\input{Commands}

\begin{document}%

\begin{titlepage}
\PHyear{2020}
\PHnumber{024}      
\PHdate{03 March}  
%

\title{Multiplicity dependence of \pion, K, and p production in pp collisions at $\sqrt{s} = 13$~TeV}
\ShortTitle{}   

\Collaboration{ALICE Collaboration\thanks{See Appendix~\ref{app:collab} for the list of collaboration members}}
\ShortAuthor{ALICE Collaboration} 

\begin{abstract}
\input{Inputs/Abstract}
\end{abstract}
\end{titlepage}
\setcounter{page}{2}

\input{PiKPvsMulti13TeV_body}

\newenvironment{acknowledgement}{\relax}{\relax}
\begin{acknowledgement}
\section*{Acknowledgements}
\input{fa_2020-02-05.tex}    
\end{acknowledgement}

\bibliographystyle{utphys}   
\bibliography{biblio}

\newpage
\appendix
\section{The ALICE Collaboration}
\label{app:collab}
\input{2020-02-05-Alice_Authorlist_2020-02-05.tex}  
\end{document}

%% file: Commands.tex
\newcommand\fig[1]{Fig.~\ref{fig:#1}\xspace}
\newcommand\pion{\ensuremath{\pi}\xspace}
\newcommand\kaon{\ensuremath{\textrm{K}}\xspace}
\newcommand\pr{\ensuremath{\textrm{p}}\xspace}
\newcommand\dEdx{\ensuremath{\td E/\td x}\xspace}

\newcommand\GeVc{\ensuremath{\textrm{GeV}/c}\xspace}
\newcommand\MeVc{\ensuremath{\textrm{MeV}/c}\xspace}
\newcommand\mean[1]{\ensuremath{\langle#1\rangle}\xspace}
\newcommand\pT{\ensuremath{p_{\textrm{T}}}\xspace}
\newcommand\pt{\ensuremath{p_{\textrm{T}}}\xspace}
\newcommand\mpT{\mean{\pT}}

\newcommand\td{\ensuremath{\textrm{d}}}
\newcommand\dNps[2]{\ensuremath{\td^{#1}N_{#2}}\xspace}
\newcommand\dEta{\ensuremath{\textrm{d}\eta}\xspace}
\newcommand\dNchdEta{\ensuremath{\dNps{}{\textrm{ch}}/\dEta}\xspace}
\newcommand\mdNde{\mean{\dNchdEta}}
\newcommand\cme[1]{\ensuremath{\sqrt{s}\ =\ #1\ \textrm{TeV}}}
\newcommand\cmenn[1]{\ensuremath{\sqrt{s_{\rm NN}}\ =\ #1\ \textrm{TeV}}}
\newcommand\mbeta{\mean{\beta_{\textrm{T}}}}
\newcommand\Tkin{\ensuremath{T_{\textrm{kin}}}\xspace}
\newcommand\dNdy{\ensuremath{\dNps{}{}/\td y}\xspace}
\newcommand{\pip}          {\ensuremath{\pi^{+}}}
\newcommand{\pim}          {\ensuremath{\pi^{-}}}
\newcommand{\pipm}          {\ensuremath{\pi^{\pm}}}
\newcommand{\kap}          {\ensuremath{\mathrm{K}^{+}}}
\newcommand{\kam}          {\ensuremath{\mathrm{K}^{-}}\xspace}
\newcommand{\kapm}          {\ensuremath{\mathrm{K}^{\pm}}}
\newcommand{\p}               {$\rm p$}
\newcommand{\pbar}         {$\rm\overline{p}$\xspace}
\newcommand{\MeV}{\ensuremath{\mathrm{MeV}}}
\newcommand\cmes{\ensuremath{\sqrt{s}}\xspace}
\newcommand\tab[1]{Table~\ref{tab:#1}\xspace}
\newcommand{\pythiaeight}{P\protect\scalebox{0.8}{YTHIA}8\xspace}
\newcommand{\specialcell}[2][c]{%
  \begin{tabular}[#1]{@{}c@{}}#2\end{tabular}}

%% file: Inputs/Abstract.tex
This paper presents the measurements of \pipm, \kapm, \pr and \pbar\ transverse momentum (\pT) spectra as a function of charged-particle multiplicity density in proton-proton (pp) collisions at \cme{13} with the ALICE detector at the LHC. Such study allows us to isolate the center-of-mass energy dependence of light-flavour particle production. The measurements reported here cover a \pT range from 0.1 \GeVc\ to 20 \GeVc\ and are done in the rapidity interval $|y|<0.5$. The \pT-differential particle ratios exhibit an evolution with multiplicity, similar to that observed in pp collisions at \cme{7}, which is qualitatively described by some of the hydrodynamical and pQCD-inspired models discussed in this paper. Furthermore, the \pT-integrated hadron-to-pion yield ratios measured in pp collisions at two different center-of-mass energies are consistent when compared at similar multiplicities. This also extends to strange and multistrange hadrons, suggesting that, at LHC energies, particle hadrochemistry scales with particle multiplicity the same way under different collision energies and colliding systems.

%% file: PiKPvsMulti13TeV_body.tex
\section{Introduction}
\input{Inputs/Introduction.tex}

\input{Inputs/Analysis.tex}

\subsection{Corrections and normalization}
\input{Inputs/Data_Analyses/CorrAndNorm.tex}
\subsection{Systematic uncertainties}
\input{Inputs/Data_Analyses/Systematics.tex}

\section{Results and discussion}
\input{Inputs/Results.tex}


\section{Summary}
\input{Inputs/Summary.tex}

%% file: Inputs/Introduction.tex
The unprecedented energies available at the Large Hadron Collider (LHC) provide unique opportunities to investigate the properties of strongly-interacting matter. Particle production at large transverse momenta (\pT) is well-described by perturbative Quantum Chromodynamics (pQCD). The soft regime (\pT $\lesssim$ 2 GeV/$c$), in which several collective phenomena are observed in proton-proton (pp), proton-lead (p--Pb), and heavy-ion (A--A) collisions, is not calculable from first principles of QCD. Instead, in order to describe bulk particle production in A--A collisions, one usually relies on hydrodynamic and thermodynamic modelling, which assumes the system to be in kinetic and chemical equilibrium~\cite{deSouza:2015ena, Andronic:2017pug}. On the other hand, the description of low-\pT particle spectra in smaller systems such as pp collisions is often based on phenomenological modelling of multi-partonic interactions (MPI) and color reconnection (CR)~\cite{Ortiz:2013yxa, Sjostrand:2014zea} or overlapping strings~\cite{Bierlich:2014xba}.

Recent reports on the enhancement of (multi)strange hadrons~\cite{Adam:2015vsf}, double-ridge structure~\cite{CMS:2012qk, Abelev:2012ola}, non-zero $v_{\textrm{2}}$ coefficients~\cite{Abelev:2014mda}, mass ordering in hadron \pT spectra, and characteristic modifications of baryon-to-meson ratios~\cite{Abelev:2013haa} suggest that collective phenomena are present at the LHC energies also in p--Pb collisions. This is further extended to even smaller systems, such as pp collisions at \cme{7}, where similar observations have been reported in high multiplicity events, indicating that the collective effects are not characteristic of heavy-ion collisions only. Furthermore, a continuous transition of light-flavor hadron-to-pion ratios as a function of charged-particle multiplicity density \dNchdEta from pp to p--Pb and then to Pb--Pb collisions was found~\cite{Acharya:2018orn,ALICE:2017jyt,Aad:2015gqa}. The observed similarities suggest the existence of a common underlying mechanism determining the chemical composition of particles produced in these three collision systems.

Results from pp~\cite{Acharya:2018orn} and p--Pb~\cite{Abelev:2013haa} collisions indicate that particle production scales with \dNchdEta independent of the colliding system. Measurements reported in previous multiplicity-dependent studies have considered different colliding systems, each at a different center-of-mass energy. In this work, we extend the existing observations by performing a detailed study of pp collisions at \cme{13}. A similar study has been reported by the CMS Collaboration, albeit in a limited \pt range~\cite{Sirunyan:2017zmn}.
Thanks to the availability of Run~2 data from the LHC, for the first time, in pp collisions, we can disentangle the effect of center-of-mass energy from the multiplicity dependence of \pipm, \kapm and \p\,(\pbar) production in a wide \pT range.

In this paper, we report on the multiplicity dependence of the production of primary \pipm, \kapm and \pr (\pbar) at \cme{13}. Particles are considered as primary if their mean proper decay length $c\tau$ is larger than 1~cm and they are created in the collision (including products of strong and electromagnetic decays), but not from a weak decay of other light-flavor hadrons or muons. An exception to this are products of weak decays, where $c\tau$ of the weakly decaying particle is less than 1~cm~\cite{ALICE-PUBLIC-2017-005}. The reported particle spectra are measured in the rapidity region $|y|<0.5$ with the ALICE detector~\cite{Aamodt:2008zz}, which offers excellent tracking and  particle identification capabilities from $\pT=0.1$ \GeVc to several tens of \GeVc~\cite{Abelev:2014ffa}. As particles and anti-particles are produced roughly in equal amounts at LHC energies~\cite{Abbas:2013rua}, we adopt a notation where \pion, \kaon, and \pr refer to $(\pip + \pim)$, $(\kap + \kam)$, and $(\pr$ + \pbar) unless stated otherwise. This paper is organized as follows. In Sec.~2, the details on particle identification techniques, systematic uncertainties, spectra corrections and normalization are provided. The results are presented and discussed in Sec.~3, together with comparisons to Monte Carlo model predictions. Finally, the most important findings are summarized in Sec.~4.

%% file: Inputs/Analysis.tex
\section{Data set and experimental setup}
The dataset used for this study was recorded by the ALICE Experiment during the 2016 LHC pp data taking period. Overall $\sim$143M events have been analysed, corresponding to an integrated luminosity of $2.47\textrm{\ nb}^{-1}$ considering the visible cross-section measured with the V0 detector~\cite{ALICE-PUBLIC-2016-002}. A detailed description of the ALICE detector and its performance is provided in~\cite{Aamodt:2008zz, Abelev:2014ffa}. Measurements of identified particle spectra have been performed by using the central barrel detectors: the Inner Tracking System (ITS) (Sec. 3.1 of~\cite{Aamodt:2008zz}), the Time Projection Chamber (TPC)~\cite{Alme:2010ke} and the Time-of-Flight detector (TOF)~\cite{Gruttola:2014kwa}. The charged-particle multiplicity estimation is done by the V0 detector (Sec. 5.4 of~\cite{Aamodt:2008zz}), which consists of two arrays of 32 scintillators each, positioned in the forward (V0A, $2.8 < \eta < 5.1$) and backward (V0C, $-3.7 < \eta < -1.7$) rapidity regions. In addition, the V0 is also used for triggering purposes as well as background rejection. The determination of the event collision time~\cite{Adam:2016ilk} is performed by the T0 detector as well as the TOF detector. The former consists of two arrays of Cherenkov counters, positioned on both sides of the interaction region, and covering a pseudorapidity range of $-3.3 < \eta < -2.9$ (T0-C) and $4.5<\eta<5$ (T0-A). The central barrel detectors are placed inside a solenoidal magnet, which provides a field strength of 0.5\,T.

The ITS is the innermost detector and consists of six concentric cylindrical layers of high-resolution silicon detectors based on different technologies, covering pseudorapidity region $|\eta|<0.9$. The two innermost layers form the Silicon Pixel Detector (SPD), which features binary readout and is also used as a trigger detector. The Silicon Drift Detector (SDD) and the Silicon Strip Detector (SSD), which form the four outer layers of the ITS, provide the amplitude of the charge signal, which is used for particle identification through the measurement of specific energy loss at low transverse momenta ($\pT \gtrsim 100$\ MeV/$c$).

The TPC, which is the main tracking detector of the ALICE central barrel, is based on a cylindrical gaseous chamber with radial and longitudinal dimensions of $85{\rm\,cm} < r < 247{\rm\,cm}$ and $-250{\rm\,cm}<z<250{\rm\,cm}$, respectively. The TPC is read out by multi-wire proportional chambers (MWPC) with cathode pad readout, located at its endplates.
With the measurement of drift time, the TPC provides three-dimensional space-point information for each charged track in pseudorapidity range $|\eta|<0.8$ with up to 159 samples per track.
In the TPC, the identification of charged particles is based on the measurement of the specific energy loss, which in pp collisions is performed with a resolution of $5.2\%$~\cite{Abelev:2014ffa}.

The TOF is a large-area array of multigap resistive plate chambers (MRPC), formed into a $\sim 4$\ m radius cylinder around the interaction point and covering the pseudorapidity region $|\eta|<0.9$ with full-azimuth coverage. The time-of-flight is measured as the difference between the particle arrival time and the event collision time, enabling particle identification at intermediate transverse momenta, $0.5 \lesssim \pT \lesssim 4$\ \GeVc. The arrival time is measured by the MRPCs with an intrinsic time resolution of 50\ ps, while the event collision time is determined by combining the T0 detector measurement with the estimate using the particle arrival times at the TOF~\cite{Adam:2016ilk}.

\subsection{Event selection, classification and normalization}

The analysed data were recorded using a minimum-bias trigger requiring signals in both V0A and V0C scintillators in coincidence with the arrival of the proton bunches from both directions. The background events produced outside the interaction region are rejected using the correlation between the SPD clusters and the tracklets reconstructed in SPD. The out-of-bunch pileup was rejected offline using the timing information from the V0 counter. The primary vertex was reconstructed either using global tracks (reconstructed using ITS and TPC information) or SPD tracklets (reconstructed using only the SPD information) with $|z_{\textrm{vtx}}| < 10$\ cm along the beam axis. Events with in-bunch pileup were removed if a second vertex was reconstructed within $8{\rm\,mm}$ of the primary vertex in the beam direction. The typical interaction rate of pp collisions in the 2016 data taking periods was around 120 kHz while beam-gas interactions occurred at a rate of 1.2 kHz.

In the analysis presented in this paper, we consider the event class INEL$>$0 with at least one charged particle produced in the pseudorapidity region $|\eta|<1$, which corresponds to $\sim75\%$ of the total inelastic scattering cross-section~\cite{Acharya:2019kyh}. To avoid auto-correlation biases~\cite{Acharya:2018orn, Acharya:2019kyh}, the events are classified using the total charge collected in the V0 detector (V0M amplitude), which scales linearly with the total number of the corresponding charged particles in its acceptance~\cite{Abbas:2013taa}. For each event class, the corresponding mean charged-particle multiplicity density \mdNde is measured at mid-rapidity $(|\eta|<0.5)$ as summarised in~\tab{mdNdeTable}.

\input{Inputs/Tables/mdNdeVsMulti.tex}
\input{Inputs/Tables/PIDpTRanges.tex}

\subsection{Identification of charged pions, kaons and protons}
In order to measure particle spectra in a wide \pT range, several sub-analyses employing different detectors and particle identification (PID) techniques were performed and combined. As a result, the combined spectra cover transverse momenta ranges from 0.1/0.2/0.3 \GeVc to 20 \GeVc for \pion/\kaon/\pr. The \pT and (pseudo)rapidity ranges covered by each analysis for different particle species are summarized in \tab{PIDptpikp}.

At low \pT, hadron spectra were measured by the ITS stand-alone (ITSsa) analysis. The dynamic range of the analogue readout of SDD and SSD allows for \dEdx measurements of highly ionizing particles, which otherwise do not reach the outer detectors. Hadron identification in the ITS is carried out by calculating the truncated mean of \dEdx and comparing it to the expected energy loss under different mass hypotheses. The difference between measured and expected \dEdx is then estimated in terms of the standard deviation $\sigma$ and the particle mass hypothesis with the lowest score is assigned. This is feasible even for pp collisions with the highest multiplicities, as the number of charge clusters wrongly assigned to the reconstructed tracks is negligible. A detailed description of the method is provided in~\cite{Acharya:2018orn}.

Hadrons at intermediate \pT enter the fiducial volume of the TPC where they can be identified by measuring the charge generated in the gas. The truncated mean of \dEdx is calculated for the global tracks and compared to the expected energy loss under a given mass hypothesis. At low transverse momenta where the separation between different species is sufficiently large, tracks within three standard deviations from the expected \dEdx are assigned to a given hypothesis. In the regions where signals from several species overlap ($\pT < 0.4~\GeVc$ for \pion, $\pT > 0.45~\GeVc$ for \kaon, and $\pT > 0.6~\GeVc$ for \pr), \dEdx is fit with two Gaussian distributions, one to describe the signal and the other to describe the tail of the overlapping species. The fit of the overlapping species is then integrated in the signal region and subtracted from the signal~\cite{Acharya:2018orn}.

In the \pT region where the statistical unfolding of the TPC signal becomes unfeasible, particle identification is performed using the time-of-flight measurements. The results presented in this paper were obtained by combining the particle spectra estimated with two separate TOF analyses, taking into account the non-common part of the respective systematic uncertainties. In the ``TOF template fits'', the PID is based on a statistical unfolding method, where the distribution of the difference between measured and expected time-of-flight (i.e. $\Delta t$) is fitted with templates for pions, kaons and protons in each \pT and multiplicity bin~\cite{Abelev:2013vea}. An additional template is needed to take into account the background due to wrongly associated tracks with hits in the TOF detector. The template for each particle is built from data, considering the measured TOF time response function (Gaussian with an additional exponential tail for larger arrival times). The fits are repeated separately for each particle hypothesis in $|y|<0.5$. In contrast to this, in the ``TOF fits'' analysis, the velocity $\beta$ distribution is simultaneously fitted for all three particle types. For this purpose, four analytic functions, three for \pion, \kaon and \pr, and one for mismatches, are employed. The analysis is performed in two narrow pseudorapidity slices ($|\eta| < 0.2$ and $0.2<|\eta|<0.4$) and in momentum bins, which are then unfolded to transverse momenta. The corresponding rapidity interval is determined under the assumption of a flat \dNchdEta distribution in the aforementioned pseudorapidity bins~\cite{Adam:2015gka}.

Charged kaons can also be identified via the kink decay topology, where a charged particle decays into a charged and a neutral daughter (${\rm K}^{\pm}\rightarrow \mu^{\pm} \nu_{\mu}$ or ${\rm K}^{\pm}\rightarrow \pi^{\pm}\pi^{0}$). This secondary vertex where both decaying particle and the charged decay product have the same charge is reconstructed inside the ALICE TPC detector. This technique extends the charged kaon identification up to 6 GeV/$c$ on a track-by-track basis. The algorithm for selecting kaons via their kink decay is used in a fiducial volume inside the TPC corresponding to a radial distance of $120<R<210$ cm. This selection allows for an adequate number of TPC clusters to be associated with the decaying particle and its products. The track of the decaying particle is required to fulfil all the criteria of the global tracks except for the minimum number of clusters, which in this case is 30.

The topological selection of the kaon candidates and their separation from the pion decays ($\pi^{\pm}\rightarrow \mu^{\pm} \nu_{\mu}$) is based on the two-body decay kinematics. The transverse momentum of the decay product with respect to the decaying particle's direction ($q_{\rm T}$) has an upper limit of 236 MeV/$c$ for kaons and 30 MeV/$c$ for pions for the two-body decay to $\mu^{\pm} \nu_{\mu}$. Similarly, for kaons decaying to pions, this limit is 205 MeV/$c$. Thus, a selection of $q_{\textrm{T}} < 120$~MeV/$c$ rejects the majority (85\%) of pion decays. In addition, the angle between the mother and the daughter tracks is selected to be above the maximum allowed decay angle for pions and below the maximum allowed decay angle for kaons~\cite{Adam:2015qaa}. The invariant mass for the decay $\mu^{\pm} \nu_{\mu}$, $M_{\mu\nu}$ is calculated by assuming the daughter track to be a muon and the undetected track to be a neutrino. These selection criteria lead to a kaon sample with a purity of $97\%$.

The strategy employed to measure particle production in the region of the relativistic rise of the TPC was reported in~\cite{Adam:2015kca}. The \dEdx signal in the relativistic rise $(3< \beta \gamma \left(= \frac{p}{m} \right) < 1000)$ follows the functional form $\ln(\beta \gamma)$. In addition to the logarithmic growth, the separation in number of standard deviations between pions and protons, pions and kaons, and kaons and protons as a function of momentum is nearly constant, which allows identification of charged pions, kaons, and (anti)protons with a statistical deconvolution approach from $p_{\rm T} \approx 2-3~{\rm GeV}/c$ up to $p_{\rm T} = 20~{\rm GeV}/c$. In order to describe the TPC response in the relativistic rise, clean external samples of secondary particles were used to parametrize the Bethe-Bloch and resolution curves. These correspond to pions (protons) from weak decays: ${\rm K}^{0}_{S} \rightarrow  \pi^{+}+\pi^{-}\ (\Lambda \rightarrow  \rm{p}+\pi^{-} )$ and electrons from photon conversion. Moreover, primary pions measured with the TOF detector were used. The parametrization is done as a function of pseudorapidity. For short (long) tracks, i.e tracks within $|\eta|<0.2\,(0.6<|\eta|<0.8)$, the resolution for protons is $\approx 6.2\%\ (\approx 5.4\%)$, while for pions it is $\approx 5.4\%\ (\approx 5.0\%)$. To extract the fraction of charged pions, kaons, and protons in the four different pseudorapidity intervals ($|\eta|<0.2$, $0.2 < |\eta| < 0.4$, $0.4 < |\eta| < 0.6$, and $0.6 < |\eta| < 0.8$) a 4-Gaussian fit (three for \pion, \kaon, \pr and one to remove the unwanted electron contribution) to the \dEdx distribution in momentum bins is performed. The only free parameter in each of the Gaussian functions is the normalization, while the $\langle {\rm d}E/{\rm d}x \rangle$ and $\sigma_{\langle {\rm d}E/{\rm d}x \rangle}$ are obtained and fixed using the Bethe-Bloch and resolution parametrizations, respectively. A weighted average of the four different measurements is calculated to obtain the particle fractions in $|\eta|<0.8$. The yields are obtained by multiplying the particle fractions by the measured unidentified charged particle spectrum.

%% file: Inputs/Tables/mdNdeVsMulti.tex
\begin{table*}[!h]
  \begin{center}
\caption{Mean charged-particle multiplicity density \mdNde measured in different event multiplicity classes. Multiplicity classes are selected based on the visible inelastic scattering cross-section. The fraction of the total inelastic scattering cross-section is quoted for each class in the central rows.}
\label{tab:mdNdeTable}
  \begin{tabular}{c c c c c c c c c c c}
    \hline \hline
    V0M mult. class & I & II & III & IV & V \\
    $\sigma$/$\sigma_{\rm INEL>0}$ (\%) & 0--0.92 & 0.92--4.6 & 4.6--9.2 & 9.2--13.8 & 13.8--18.4  \\
    \mdNde  & $26.02\pm0.35$ & $20.02\pm0.27$ & $16.17\pm0.22$ & $13.77\pm0.19$ & $12.04\pm0.17$  \\ \midrule
    V0M mult. class & VI & VII & VIII & IX & X\\
    $\sigma$/$\sigma_{\rm INEL>0}$ (\%) & 18.4--27.6 & 27.6--36.8 & 36.8--46.0 & 46.0--64.5 & 64.5--100 \\
    \mdNde  & $10.02\pm0.14$ & $7.95\pm0.11$ & $6.32\pm0.09$ & $4.50\pm0.07$ & $2.55\pm0.04$ \\  \bottomrule
  \end{tabular}
  \end{center}
\end{table*}

%% file: Inputs/Tables/PIDpTRanges.tex
\begin{table*}[!h]
\label{PIDpt}
  \caption{Different \pt ranges used for the identification of pions, kaons and protons. The final \pt spectra have been obtained by combining the results of the various PID techniques.}
  \label{tab:PIDptpikp}
  \begin{tabularx}{\textwidth}{p{3.0cm}*{2}{Y}*{2}{Y}*{2}{Y}*{4}{Y}}
\hline\hline

 \multirow{2}{*}{Analysis} & \multirow{2}{*}{PID Technique} & \multicolumn{6}{c}{\pt ranges (\GeVc)} & (pseudo)rapidity\\
    &  & \multicolumn{2}{c}{\pipm} & \multicolumn{2}{c}{\kapm} & \multicolumn{2}{c}{\p\ (\pbar)} & range \\
    \hline \\

	ITSsa & n$\sigma$ integral  & \multicolumn{2}{c}{0.1$-$0.7} & \multicolumn{2}{c}{0.2$-$0.6} & \multicolumn{2}{c}{0.3$-$0.65} & $|y|<$0.5 \\ \\

	TPC-TOF fits & \specialcell{n$\sigma$ fits to TPC,\\ $\beta$ fits to TOF} & \multicolumn{2}{c}{0.25$-$3.0} & \multicolumn{2}{c}{0.3$-$3.0} & \multicolumn{2}{c}{0.45$-$3.0} & \specialcell{$|y|<$0.5 (TPC)\\ $|\eta|<$0.4 (TOF)} \\ \\

	TOF template fits & \specialcell{Statistical\\ unfolding of $\Delta t$} & \multicolumn{2}{c}{0.7$-$4.0} & \multicolumn{2}{c}{0.6$-$3.0} & \multicolumn{2}{c}{0.9$-$4.0} & $|y|<$0.5 \\ \\

	Kinks & Kink topology & \multicolumn{2}{c}{$-$} & \multicolumn{2}{c}{0.35$-$6.0} & \multicolumn{2}{c}{$-$} & $|y|<$0.5 \\ \\

	rTPC & TPC d\textit{E}/d\textit{x} fits & \multicolumn{2}{c}{2$-$20} & \multicolumn{2}{c}{3$-$20} &  \multicolumn{2}{c}{3$-$20} & $|\eta|<$0.8 \\ \\

\hline \hline
	\end{tabularx}

\end{table*}

%% file: Inputs/Data_Analyses/CorrAndNorm.tex
The raw particle distributions are normalized to the total number of events analysed\footnote{Events that passed all the selection criteria.} in each multiplicity class. To obtain the \pT distributions of primary \pion, \kaon, and \pr, the raw particle distributions obtained from the different PID approaches need to be also corrected for the detector efficiency and acceptance, the ITS-TPC, and TPC-TOF matching efficiency, the PID efficiency, the trigger efficiency and the contamination from secondary particles.\\
Secondary particles are either produced in weak decays or from the interaction of particles with the detector material.
The estimation of secondary particle contribution is based on the Monte Carlo (MC) templates of the distance of closest approach of the track to the primary vertex in the transverse plane with respect to the beam axis (DCA$_{xy}$), as carried out in previous works~\cite{Abelev:2013vea, Acharya:2018orn}. The DCA$_{xy}$ distributions of the tracks in data are fitted with three MC templates corresponding to the expected shapes of primary particles, secondaries from material and secondaries from weak decays to obtain the correct fraction of primary particles in the data. This procedure is repeated in each \pT and multiplicity bin and thus takes into account the possible differences in the feed-down corrections due to the change in the abundances and spectral shapes of the weakly decaying particles. The contamination is different in each PID analysis due to different track selection criteria and PID techniques and hence it is estimated separately for each analysis. The contribution of secondary particles was found to be significant for \pion (up to 2\%) and \pr (up to 15\%) whereas the contribution for \kaon is negligible. \\
The spectra are corrected for the detector acceptance and track reconstruction efficiencies based on a simulation using the \pythiaeight (Monash-2013 tune) Monte Carlo event generator \cite{Skands:2014pea} and particle propagation through the full ALICE geometry using GEANT3~\cite{Brun:1119728}. In this simulation, tracks are reconstructed using the same algorithms as for the data. The detector acceptance and reconstruction efficiencies are found to be independent of charged-particle multiplicity and thus the multiplicity-integrated values are used in all multiplicity classes.
As GEANT3 does not fully describe the interaction of low-momentum \pbar and \kam with the detector material, an additional correction factor to the efficiency for these two particles is estimated with GEANT4~\cite{Agostinelli:2002hh} and FLUKA~\cite{Ferrari:2005zk}, respectively, where the interaction processes are known to be better reproduced \cite{Abelev:2013vea}. Additional corrections to the efficiency are applied when TPC or TOF information is used to take into account the track matching between ITS and TPC, and between TPC and TOF.
\\
Signal losses due to the trigger selection are extracted from \pythiaeight (Monash-2013 tune) MC simulation as performed in~\cite{Acharya:2019kyh}.
The correction is found to be 17--18\% at low \pT in the V0M class X (the lowest multiplicity), and reduces to $\sim$5\%, $\sim$2\% in classes IX and VIII, respectively. The correction is negligible in higher multiplicity pp collisions and for \pT $\gtrsim$ 4 GeV/$c$ in all multiplicity bins except in class X. In the latter, the correction reaches $\sim$2\% at \pT = 7 GeV/$c$. 
Finally, an additional correction is applied to pass from triggered INEL$>$0 to true INEL$>0$ events, i.e. events with at least one primary charged particle in $|\eta^{\rm true}|$ $<$ 1 and with the primary vertex in the region $|V^{\rm true}_{\rm z}|$ $<$ 10 cm. The correction is independent of particle species and is found to be negligible from V0M I (the highest multiplicity) to V0M VI, while it ranges from 1\% in class VII to 11\% in class X. The correction is about 8\% for multiplicity-integrated INEL$>0$ events.

%% file: Inputs/Data_Analyses/Systematics.tex
The systematic uncertainties are divided into two categories, those common to all analyses and those which are analysis specific.
The common systematic uncertainties are those due to tracking, which includes track quality criteria and the \pt-dependent ITS-TPC matching efficiency (except for the ITSsa analysis), the TPC-TOF matching efficiency (for TPC-TOF and TOF analyses), and the signal loss correction. In addition, the systematic uncertainty related to the effect of the material budget on the global tracking (\pT dependent) is also added. The uncertainties on global tracking and TPC-TOF matching due to material budget are calculated by varying the material budget in the simulation by $\pm$ 5\%. The uncertainty related to the hadronic interaction cross section in the detector material is estimated using GEANT4~\cite{Agostinelli:2002hh} and FLUKA~\cite{Ferrari:2005zk} transport codes. Finally, an additional systematic uncertainty of 2\% is added to account for possible multiplicity dependence of track reconstruction efficiency and signal loss correction calculated from a MC simulation.
All common sources of systematic uncertainties are summarised in \tab{syspikp}.
In the same table, the individual analysis systematic uncertainties are also listed for each particle species.

The estimation of the systematic uncertainties for the ITSsa analysis is described in detail in \cite{Abelev:2013vea, Acharya:2018orn}.
The ITSsa tracking uncertainties are estimated by varying the main criteria for the track selection, namely those on the DCA$_{xy}$, on the $\chi^{2}$ of the track, and on the number of clusters required in the ITS layers.
The uncertainty related to the particle identification is calculated by using a Bayesian technique and comparing the results obtained with the standard n$\sigma$ method as already performed in \cite{Adam:2016dau}.
Due to the Lorentz force, the positions of ITS clusters are shifted depending on the magnetic field polarity, giving rise to a 3\% uncertainty.
Finally, the energy-independent uncertainty related to the ITS material budget is estimated with a simulation of pp collisions at $\sqrt{s}$ = 900 GeV by varying the material budget of the ITS by $\pm$7.5\%  \cite{Aamodt:2011zj}.
For the TPC-TOF fits analysis at low \pT (below 500, 600, and 800 \MeVc for \pion, \kaon, and \pr, respectively), the systematic uncertainty associated with the PID technique is calculated by integrating the measured $\textrm{d}E/\textrm{d}x$ of charged tracks in the ranges of $\pm3.5\sigma$ and $\pm2.5\sigma$, where $\sigma$ represents one standard deviation from the $\langle\textrm{d}E/\textrm{d}x\rangle$ under given mass hypothesis.
At higher \pT values, where only the time-of-flight information is used, the associated uncertainties are calculated by simultaneously varying the width and tail parameters by 10\%. An additional uncertainty is calculated by fixing the central values of the fit functions to the $\beta$ calculated for each particle species in a given momentum range. This was found to be the dominant source of systematic uncertainty for \pion and \kaon at the highest \pT values ($\gtrsim 2.5~\GeVc$).
For the TOF template fits analysis, PID uncertainties are estimated by simultaneously varying the spread and tail slope of each $\Delta t$ template by 10\%. In addition to this, for both the TPC-TOF and TOF template fits analyses, systematic uncertainties associated with tracking are calculated by varying the track selection criteria: the number of crossed rows in the TPC, the distance of closest approach in beam and transverse directions, and the quality of the global track fit $\chi^{2}$.
For the kink analysis the sources of systematic uncertainties are: the kink vertex finding efficiency (3\% constant in \pt), the kink PID efficiency (calculated by taking into account the position of the kink vertex, the number of TPC clusters of the decaying particle track, and the $q_{\rm T}$ of the decay product), and the uncertainty related to the purity of the selected sample.
The contamination due to the random association of tracks wrongly attributed to kaon decays is of the order of 2.3\% at low transverse momenta and reaches the value of 3.4\% above 4 GeV/$c$.
The largest component of the systematic uncertainties in the analysis of the relativistic rise of the TPC arises from the imprecise parametrization of both the Bethe-Bloch and resolution curves.
To quantify this uncertainty, the variations of the Bethe-Bloch resolution parametrizations with respect to the measured $\langle {\rm d}E/{\rm d}x \rangle(\sigma_{\langle {\rm d}E/{\rm d}x \rangle})$ are used to vary the values of the mean and $\sigma$ in the 4-Gaussian fit~\cite{Adam:2015kca}.
The largest relative deviation between the nominal particle ratios and the ones obtained after the variations are assigned as a systematic uncertainty.

\input{Inputs/Tables/SpectraSystTable.tex}

%% file: Inputs/Tables/SpectraSystTable.tex
\begin{table*}[ph!]\label{syst}
\caption{Sources of the relative systematic uncertainties of the \pt-differential yields of \pion, \kaon,\ and \pr. The uncertainties are split into two categories, the common and the individual-analysis specific for low, intermediate and high \pt. Numbers in parenthesis in the \pr column refer to \pbar uncertainties. In the last rows, the maximum (among multiplicity classes) total systematic uncertainty is reported.}
\label{tab:syspikp}
  \begin{tabularx}{\textwidth}{p{5.3cm}*{2}{Y}*{1}{Y | }*{2}{Y}*{1}{Y | }*{3}{Y}}
\hline\hline
    &  \multicolumn{9}{c}{Uncertainty (\%)} \\
    {\bf Common source}  & \multicolumn{3}{c}{\pion} & \multicolumn{3}{c}{\kaon} & \multicolumn{3}{c}{\pr (\pbar)} \\
\hline
    \pt\ (\GeVc)   &   0.1 & 3.0 & 20.0   &   0.2 & 2.5 & 20.0 &   0.3 & 4.0 & 20.0   \\
    \hline
    Correction for secondaries & 1  & 1 & \multicolumn{1}{c | }{1} & \multicolumn{3}{c | }{negl.} & 4 & 1 & 1 \\
    Hadronic interactions & 2 & 2.4 & 2.4 & 2.7 & 1.8 & 1.8 & 1\ (3.6) & 1\ (3.6) & 1\ (3.6) \\
    ITS-TPC matching efficiency  & 0.7 & 1.5 & 2.9 & 0.7 & 1.5 & 2.9 & 0.7 & 1.5 & 2.9\\
   Global tracking efficiency & \multicolumn{3}{c | }{0.7} & \multicolumn{3}{c | }{0.5} & \multicolumn{3}{c}{1.5} \\
   TOF matching efficiency \newline (TPC-TOF fits,TOF template fits) & \multicolumn{3}{c | }{\multirow{2}{*}{3}} & \multicolumn{3}{c | }{\multirow{2}{*}{6}} & \multicolumn{3}{c}{\multirow{2}{*}{4}} \\
   Signal-loss correction & \multicolumn{3}{c | }{0.2} & \multicolumn{3}{c | }{1} & \multicolumn{3}{c}{3.3} \\
   \hline
    \pt\ (\GeVc)   &   0.3 & 3.0 & 20.0   &   0.3 & 2.5 & 20.0 &   0.4 & 4.0 & 20.0   \\
    \hline
	Material budget \newline (TPC-TOF) & \multirow{2}{*}{0.5} & \multirow{2}{*}{1.0} & \multirow{2}{*}{0.2} & \multirow{2}{*}{1.5} & \multirow{2}{*}{1.0} & \multirow{2}{*}{0.4} & \multirow{2}{*}{2.9} & \multirow{2}{*}{1.7} & \multirow{2}{*}{0.1} \\
\hline
    {\bf Specific source} & \multicolumn{3}{c}{\pion} & \multicolumn{3}{c}{\kaon} & \multicolumn{3}{c}{\pr (\pbar)} \\
\hline
    {\bf ITSsa}, \pt\ (\GeVc)   &   0.10 & & 0.70   &   0.20 & & 0.60 &   0.30 & & 0.65   \\
    \hline
    Tracking & 1.4 &  & 1.4 & 1.5 &  & 1.5 & 1.1 &  & 1.1  \\
    Material budget & 4.8 & & 0.3 & 2.3 & & 0.6 & 5.0 & & 0.9 \\
    $E \times B$ effect & & 3.0 & & & 3.0 & & & 3.0 & \\
    PID & 0.4 & & 0.4 & 3.9 & & 3.9 & 4.2 & & 4.2 \\
\hline\hline


    {\bf TOF templates}, \pt\ (\GeVc)   &   0.7 &  & 4.0   &   0.6 & & 3.0  &   0.9 & & 4.0    \\
    \hline
    PID & 1 & & 9.4 & 1 & & 12 & 1 & & 21  \\
\hline\hline

    {\bf TPC-TOF fits}, \pt\ (\GeVc)   &   0.3 &  & 3   &   0.3 & & 3  &   0.4 & & 3    \\
    \hline
    PID & 1.4 & & 7 & 3 & & 16 & 1 & & 4.3 \\
\hline\hline

    {\bf rTPC}, \pt\ (\GeVc)   &  2.0 &  & 20.0   &   2.0 & & 20.0  &   2.0 & & 20.0    \\
    \hline
    Bethe--Bloch parameterization & 8.1 &  & 4 & 14.8 &  & 8.0  & 14 & & 6.0 \\
    Feed-Down  & 0.5 &  & 0.5 & - &  & -  & 2.4 & & 2.0 \\
\hline\hline

    {\bf Kinks}, \pt\ (\GeVc)   &  0.6 &  & 6.0   &   0.6 & & 6.0  &   0.6 & & 6.0    \\
    \hline
    PID & - & & - & 0.75 & & 5.3  & - & & - \\
    Kink vertex finding efficiency & - & & - & 3 & & 3  & - & & - \\
    Contamination & - & & - & 2.3 & & 3.4  & - & & - \\
\hline\hline

   {\bf Total}  & \multicolumn{3}{c}{\pion} & \multicolumn{3}{c}{\kaon} & \multicolumn{3}{c}{\pr (\pbar)} \\
\hline
    \pt\ (\GeVc)   &   0.3 & 3.0 & 20.0   &   0.3 & 3.0 & 20.0 &   0.3 & 3.0 & 20.0   \\
    \hline
    Total & 6.6 & 5.1 & 4.3 & 6.6 & 6.8 & 7.8 & 9.6 & 9.7 & 12.4 \\
    \hline\hline

       {\bf Particle ratios}  & \multicolumn{3}{c}{K/$\pi$} & \multicolumn{3}{c}{p/$\pi$}  \\
\hline
    \pt\ (\GeVc)   &   0.2 & 3.0 & 20.0   &   0.3 & 3.0 & 20.0    \\
    \hline
    Total & 7.2 & 14.6 & 7.7 & 10.2 & 12.2 & 11.5 \\
    \bottomrule
  \end{tabularx}
  ~\newline
\end{table*}

%% file: Inputs/Results.tex
\begin{figure}
        \begin{center}
        \includegraphics[width=.45\textwidth,page=1]{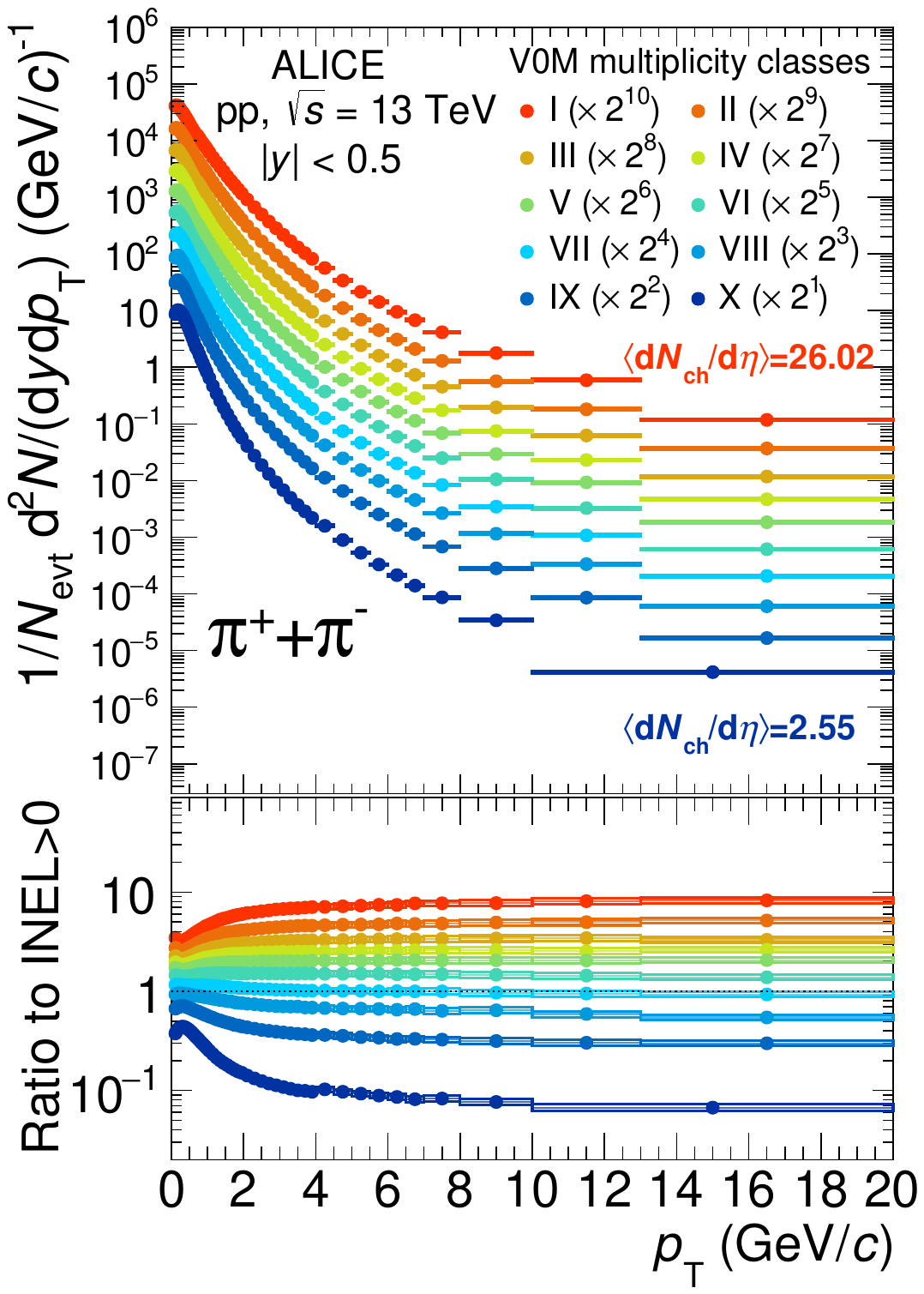}
        \includegraphics[width=.45\textwidth,page=2]{Figures/Spectra13.pdf}
        \includegraphics[width=.45\textwidth,page=3]{Figures/Spectra13.pdf}
        \caption{Transverse momentum spectra of \pion, \kaon, and \pr for different multiplicity event classes. Spectra are scaled by powers of 2 in order to improve visibility. The corresponding ratios to INEL$>$0 spectra are shown in the bottom panels.}
        \label{fig:Spectra}
        \end{center}
\end{figure}

The \pT-differential spectra of \pion, \kaon, and \pr measured as a function of the charged-particle multiplicity density in pp collisions at \cme{13} are shown in \fig{Spectra}. For each V0M class, charged-particle multiplicity density has been measured in the central region ($|\eta|<0.5$), as summarized in~\tab{mdNdeTable}. The bottom panels in \fig{Spectra} show spectral ratios to the INEL$>$0 (sum of all V0M classes) class. We observe that the measured \pT spectra become harder with increasing \mdNde, and the effect is more pronounced for protons. The hardening of the inclusive charged-hadron spectra with \mdNde has been also recently reported in~\cite{Acharya:2019mzb}, where different MPI models were shown to describe such effect. On the other hand, the mass dependence of spectral shape modifications is also observed in Pb--Pb collisions at \cmenn{2.76}~\cite{Adam:2015kca}, where it is usually associated with the hydrodynamical evolution of the system. At higher \pT ($\gtrsim 8\ \GeVc$), we find that slopes of particle spectra become independent of the multiplicity class considered, as expected from pQCD calculations~\cite{Kretzer:2000yf}.

\begin{figure}[!h]
        \begin{center}
        \includegraphics[width=\textwidth]{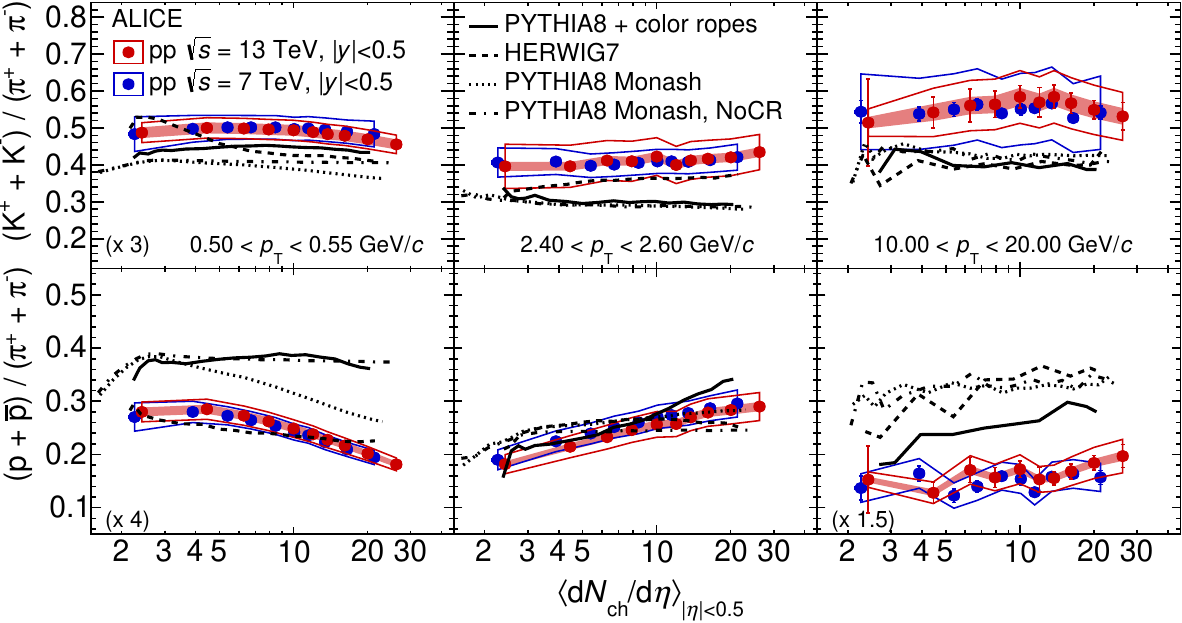}
        \caption{Multiplicity dependence of \pT-differential \kaon/\pion (upper panels) and \pr/\pion (lower panels) ratios measured in pp collisions at $\sqrt{s}$ = 7 TeV \cite{Acharya:2018orn} and 13 TeV (blue and red, respectively).
        Lines represent different MC generator predictions for pp collisions at \cme{13}. Left to right: low-, intermediate-, and high-transverse momenta. Vertical bars, open, and shaded bands represent statistical, total systematic, and multiplicity uncorrelated systematic uncertainties, respectively. Numbers in the parenthesis in different panels represent different scale factors for data and MC predictions for better readability.}
        \label{fig:pTRatios}
        \end{center}
\end{figure}

The \pT-differential \kaon/\pion and \pr/\pion ratios as a function of \mdNde measured at low, intermediate, and high transverse momenta are shown in \fig{pTRatios} together with those measured in pp collisions at \cme{7}~\cite{Acharya:2018orn} and predictions from several MC generators for pp collisions at \cme{13}.
The measured \kaon/\pion ratio shows no evident sign of evolution with multiplicity in all \pT ranges considered, while the \pr/\pion ratio shows depletion at low \pT, an increase at intermediate \pT, and constant behavior at high \pT. In addition, the measured \kaon/\pion and \pr/\pion ratios are consistent between the two center-of-mass energies~\cite{Acharya:2018orn}.

For MC predictions, the event classification is based on the number of charged tracks simulated at forward and backward pseudorapidities covered by the V0 detector, in a way similar to the way the event classification is done for the data.
The mean charged-particle multiplicity density is then calculated in the central pseudorapidity region, $|\eta|<0.5$.
HERWIG~7~\cite{Bellm:2015jjp,Bahr:2008pv}, where a clustering approach is used for hadronization, provides a good description of the evolution of the \kaon/\pion and \pr/\pion ratios with \mdNde in the low and intermediate \pT ranges and is consistent with the measured ratios within 1-2 standard deviations. \pythiaeight~\cite{Sjostrand:2007gs} without color reconnection (CR) predicts no evolution of \kaon/\pion and \pr/\pion ratios. The CR scheme, which has been shown to capture the modifications of the baryon-to-meson ratios~\cite{Ortiz:2013yxa}, provides only a qualitative description of the evolution of the \pr/\pion ratio with \mdNde and overestimates the absolute values of the ratio at low and high \pT. The implementation of color ropes~\cite{Flensburg:2011kk,Bierlich:2014xba,Bierlich:2018xfw} in \pythiaeight, which results in higher effective string tension and thus enhances strange- and di-quark production, provides a qualitative description \kaon/\pion (\pr/\pion) ratio only at low (intermediate) \pT and overestimates the \pr/\pion ratio at low \pT. This could be understood considering that larger effective string tension is mostly translated to hadronic mass and thus feeds down the low \pT part of the spectrum.

In large collision systems such as Pb--Pb, multiplicity-dependent modifications of hadron \pT spectra can be interpreted as the hydrodynamical radial expansion of the system and studied in the context of the Boltzmann-Gibbs Blast-Wave model~\cite{Schnedermann:1993ws}. In this model, a thermalized medium expands radially and undergoes an instantaneous kinematic freeze-out. The average expansion velocity \mbeta, the kinetic freeze-out temperature \Tkin, and the velocity profile exponent $n$ can be extracted from simultaneous model fits to hadron spectra. As the trends observed in the evolution of particle spectra measured in pp collisions are highly reminiscent to those in p--Pb and Pb--Pb, it is interesting to check whether the Blast-Wave model can be  extended to describe pp collisions. Such study has been previously reported in~\cite{Acharya:2018orn}, where pp, p--Pb, and Pb--Pb collisions at \cmenn{\textrm{7, 5.02, and 2.76}} were considered. Now, for the first time, we can study the evolution of \mbeta, \Tkin and $n$ in pp collisions as a function of the collision energy.

\nobreak

\begin{figure}
        \begin{center}
        \includegraphics[width=.7\textwidth]{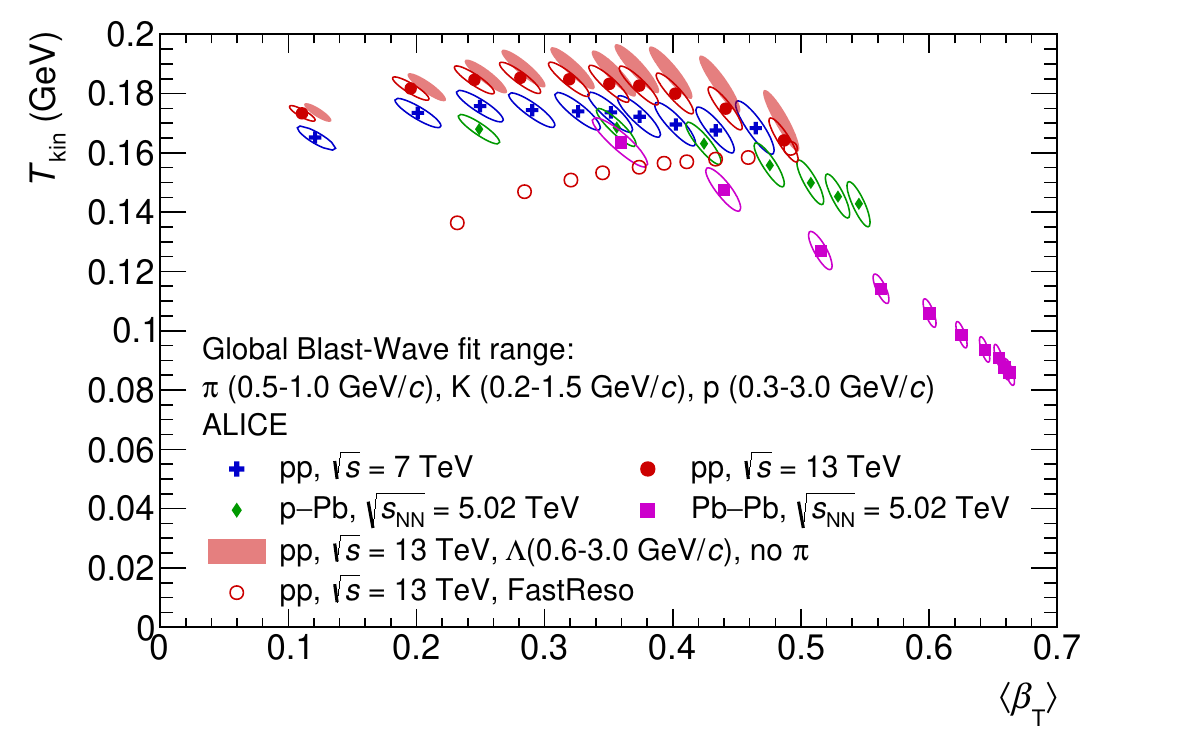}
        \caption{Correlation of kinetic freeze-out temperature \Tkin and average expansion velocity \mbeta, extracted from simultaneous Blast-Wave fits to \pion, \kaon, and \pr spectra measured in pp, p--Pb, and Pb--Pb collisions. Contours represent 1$\sigma$ uncertainty. The shaded ellipses represent the \Tkin-\mbeta correlation calculated from Blast-Wave fit to \kaon, \pr, and $\Lambda$ spectra~\cite{Acharya:2019kyh} measured in pp collisions at \cme{13}. The empty circles represent Blast-Wave fits with resonance decays~\cite{Mazeliauskas:2019ifr}. References from~\cite{Abelev:2013haa,Acharya:2018orn,Mazeliauskas:2019ifr,ALICE:2017jyt}.}
        \label{fig:BWFits}
        \end{center}
\end{figure}

\begin{figure}
        \begin{center}
        \includegraphics[width=.32\textwidth]{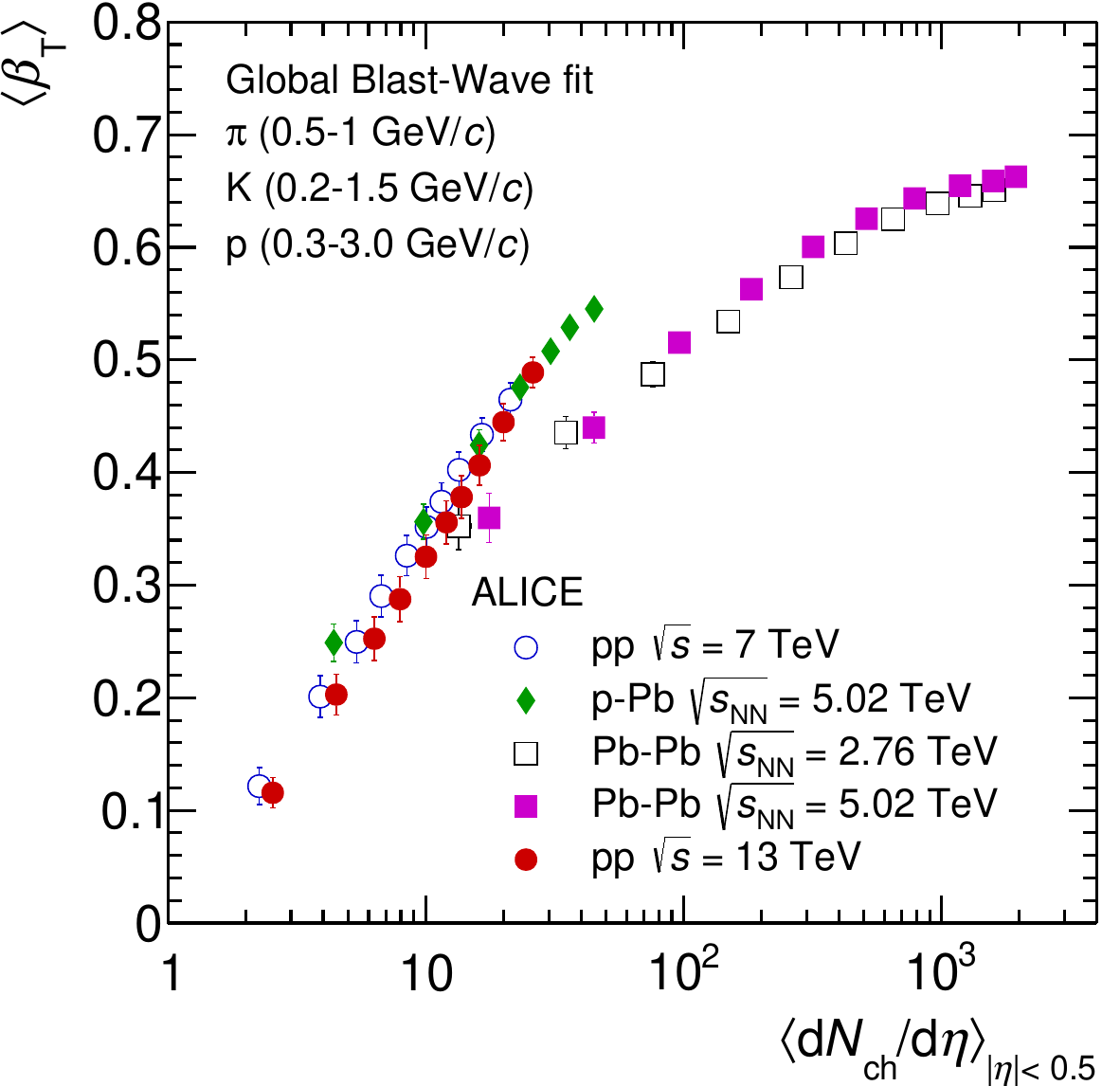}
        \includegraphics[width=.32\textwidth]{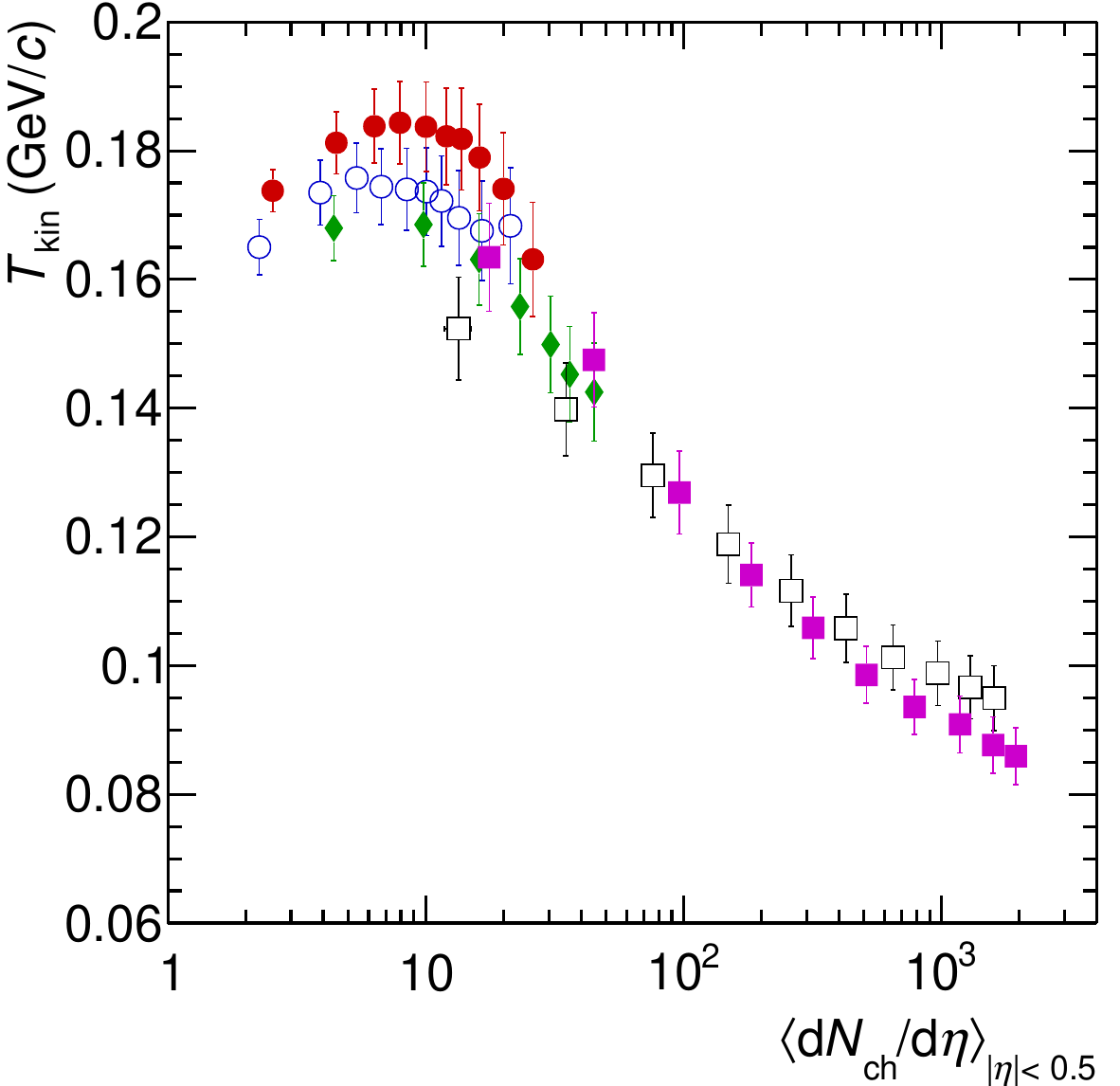}
        \includegraphics[width=.32\textwidth]{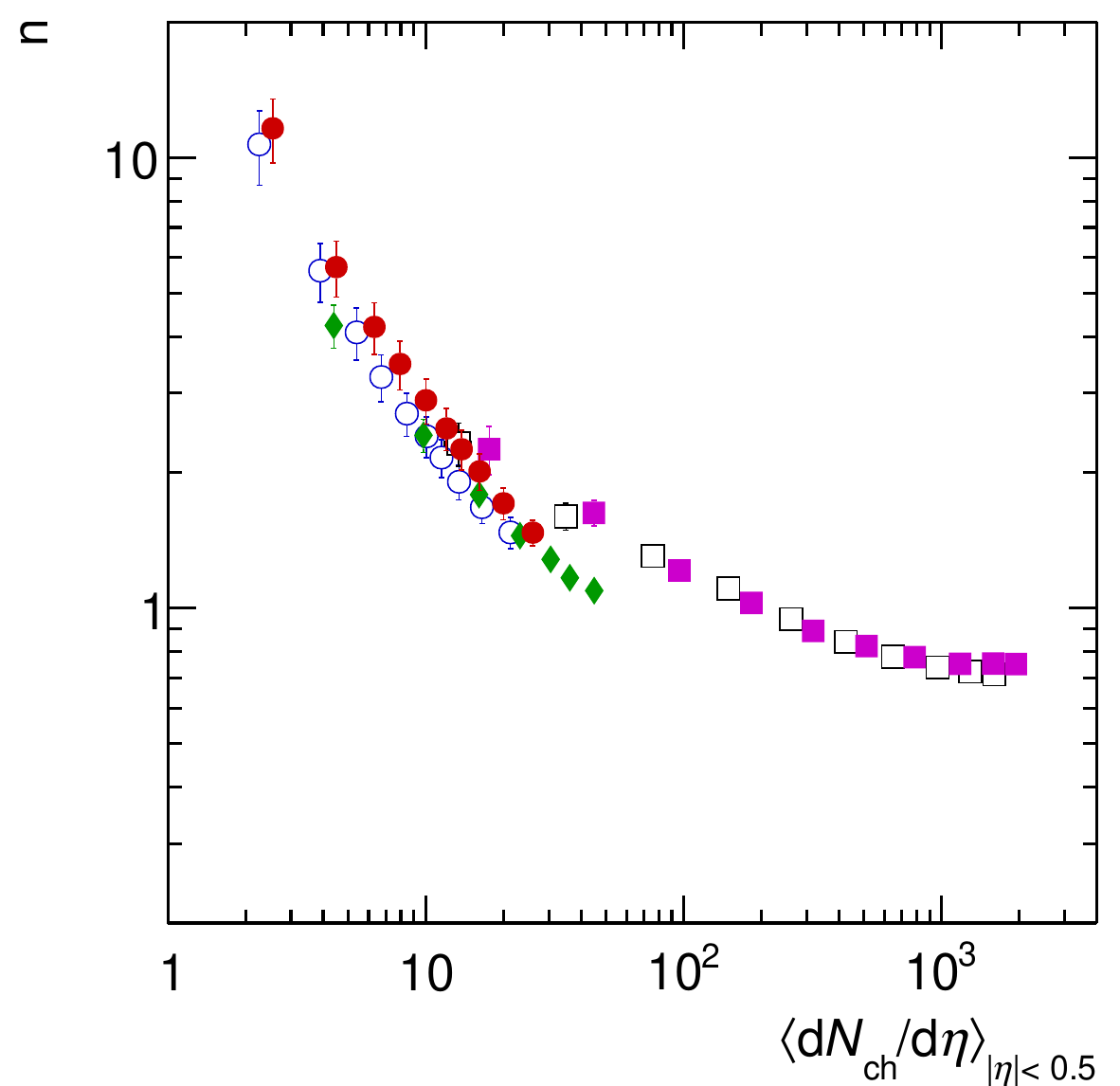}
        \caption{Evolution of \mbeta, \Tkin, and $n$ with \mdNde. \mbeta, \Tkin and $n$ are extracted from Blast-Wave fits to \pion, \kaon, and \pr \pT spectra measured in pp, p--Pb, and Pb--Pb collisions at different \cmes. The resonance feed-down contribution is neglected.}
        \label{fig:BWFitParameters}
        \end{center}
\end{figure}

At low transverse momenta ($\pT \lesssim 500~\MeVc$), the dominant mechanism of \pion production is from resonance decays. To account for this in the Blast-Wave model fits, spectral measurements of all strongly decaying hadrons are required. Alternatively, one can choose to omit the low-\pT pions. Noting that there is a strong dependence of Blast-Wave parameters on the fitting range~\cite{Abelev:2013vea}, it is important to consider the same \pT range in the fitting procedure in order to obtain a consistent comparison between different colliding systems.
The comparison of the \mbeta-\Tkin correlations measured in different systems and center-of-mass energies is shown in \fig{BWFits}.
In this paper we consider three different approaches to the Blast-Wave model fits to particle spectra measured in pp collisions at \cme{13}: a) traditional fits as done in~\cite{Abelev:2013vea, Abelev:2013haa, Acharya:2018orn}, where \pion, \kaon, and \pr spectra are fitted and resonance feed-down is neglected (represented by full markers in~\fig{BWFits}), b) simultaneously fitting \kaon, \pr, and $\Lambda$ spectra~\cite{Acharya:2019kyh} noting that $\Lambda$ are not significantly affected by resonance decays (represented by shaded ellipses in~\fig{BWFits}), and c) a method proposed in~\cite{Mazeliauskas:2018irt, Mazeliauskas:2019ifr}, where the resonance feed-down is calculated before the Cooper--Frye freeze-out using a statistical hadronization model (represented by empty circles in~\fig{BWFits}). We find that the \mbeta-\Tkin correlation in pp collisions at \cme{13} follows similar trends as seen at lower energies. When $\Lambda$'s are considered instead of pions, the trends seen in \mbeta-\Tkin correlation do not change significantly and only at highest multiplicities we find a larger \Tkin. On the other hand, when a proper treatment of resonance decays is used, we find a significantly lower \Tkin of around 135~MeV at the lowest multiplicities, which then grows with increasing \mdNde and approaches the pseudocritical QCD temperature $T_{\rm c} = 156 \pm 1.5~\MeV$~\cite{Bazavov:2018mes} at the highest multiplicity pp collisions.
In addition, the evolution of \mbeta, \Tkin, and $n$ with \mdNde is shown in \fig{BWFitParameters} for different colliding systems. From the lowest multiplicities, \Tkin grows with \mbeta until it saturates at around 180\,MeV. At larger multiplicities ($\mdNde\gtrsim 16$), \Tkin decreases and becomes similar to that measured in p--Pb collisions at \cme{5.02}, suggesting that the system decouples at lower temperature and thus is longer-lived. The average expansion velocity \mbeta increases with \mdNde and its values are consistent for pp collisions at different $\sqrt{s}$ as well as with the corresponding values for p--Pb collisions, indicating that small systems become more explosive at larger multiplicities. In contrast to this, \mbeta measured in Pb--Pb collisions is lower than that in smaller systems for the common \mdNde range, see \fig{BWFitParameters}. This indicates that the size of the colliding system might have significant effects on the final state particle dynamics. This is also reflected in the expansion velocity profile power $n$ shown in~\fig{BWFitParameters}: in pp and p--Pb collisions, large $n$ suggests high pressure gradients which lead to larger \mbeta, while in Pb--Pb collisions, $n\sim 1$ could be interpreted as lower pressure gradient and thus smaller expansion velocity~\cite{Shuryak:2013ke}.

Previous studies on hadron production as a function of multiplicity have reported the factorization of \pT-integrated particle yields with \mdNde~\cite{Acharya:2018orn}, which extends across different colliding systems and collision energies. Now for the first time we can isolate the center-of-mass energy dependence of this scaling for \pion, \kaon, and \pr in pp collisions.
The \pT-integrated particle yields (\dNdy) and average transverse momenta (\mpT) are calculated by integrating the measured transverse momentum spectra and using the L\'{e}vy-Tsallis parametrization~\cite{Tsallis:1987eu, Wilk:1999dr, Adams:2004ep} to extrapolate to the low \pT regions not covered by the measurements. The extrapolated fractions of the yields at low \pT are 8\% (10\%) for \pion, 6\% (13\%) for \kaon, and 7\% (20\%) for \pr for the highest (lowest) multiplicities. For systematic uncertainties on the extrapolation, Bylinkin, Bose-Einstein, Fermi-Dirac, $m_{\rm T}$-exponential and Hagedorn functions are used to fit particle spectra. The largest systematic uncertainties on \dNdy~(\mpT) related to the extrapolation procedure are found to be 2\% (2\%), 2\% (2\%), and 3\% (2\%) for \pion, \kaon, and \pr at low-multiplicity classes and become smaller at higher multiplicities.
\begin{figure}
  \begin{center}
    \includegraphics[width=.8\textwidth]{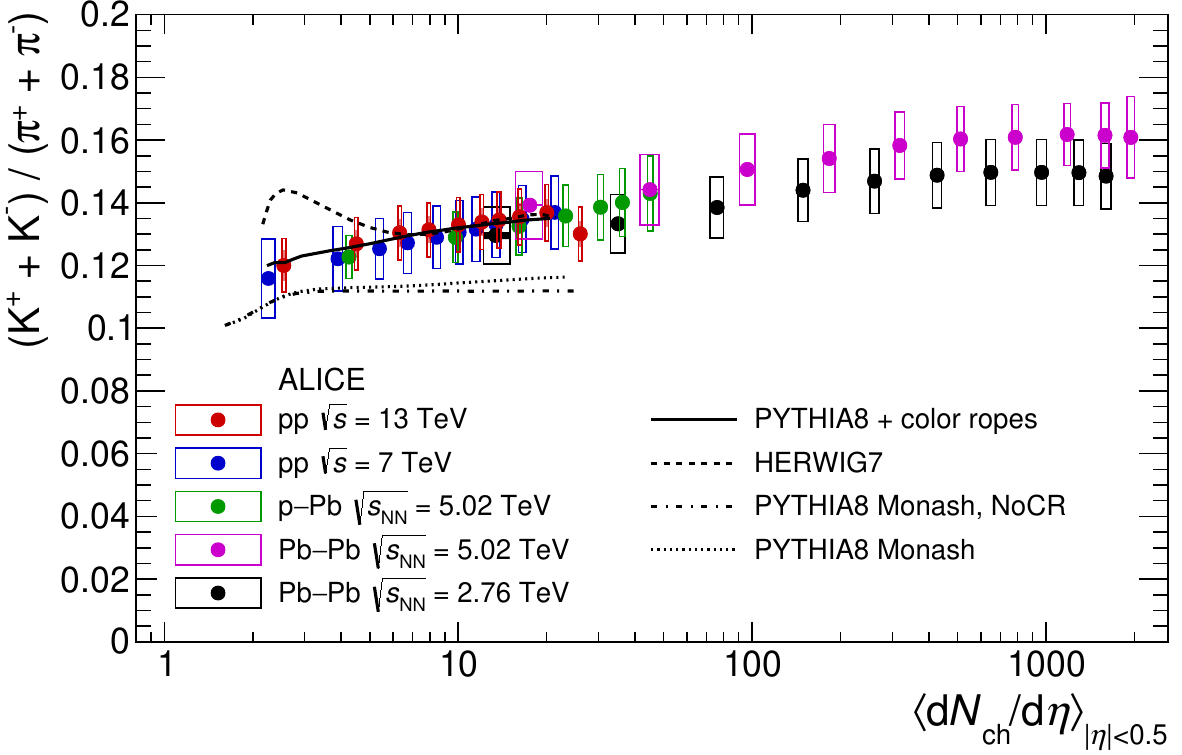}
    \includegraphics[width=.8\textwidth]{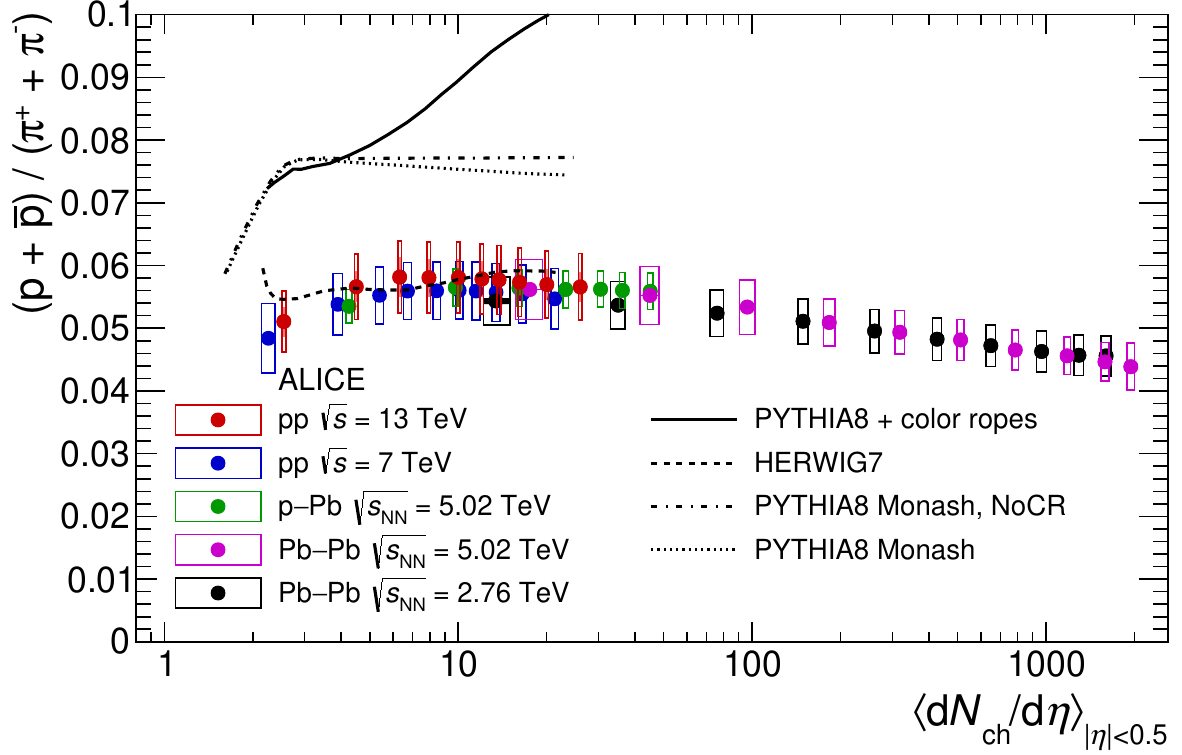}
    \caption{Integrated \kaon/\pion (top) and \pr/\pion (bottom) yield ratios as a function of charged-particle multiplicity density measured in pp, p--Pb, and Pb--Pb collisions at different center-of-mass energies. Empty (shaded) boxes represent total (multiplicity uncorrelated) systematic uncertainties. Black lines represent predictions from different MC generators for pp collisions at \cme{13}. References from~\cite{Acharya:2018orn,Adam:2016dau,Abelev:2013vea,Acharya:2019yoi}.}
    \label{fig:IntPartRatioVsMulti}
  \end{center}
\end{figure}

\begin{figure}
  \begin{center}
    \includegraphics[width=.6\textwidth]{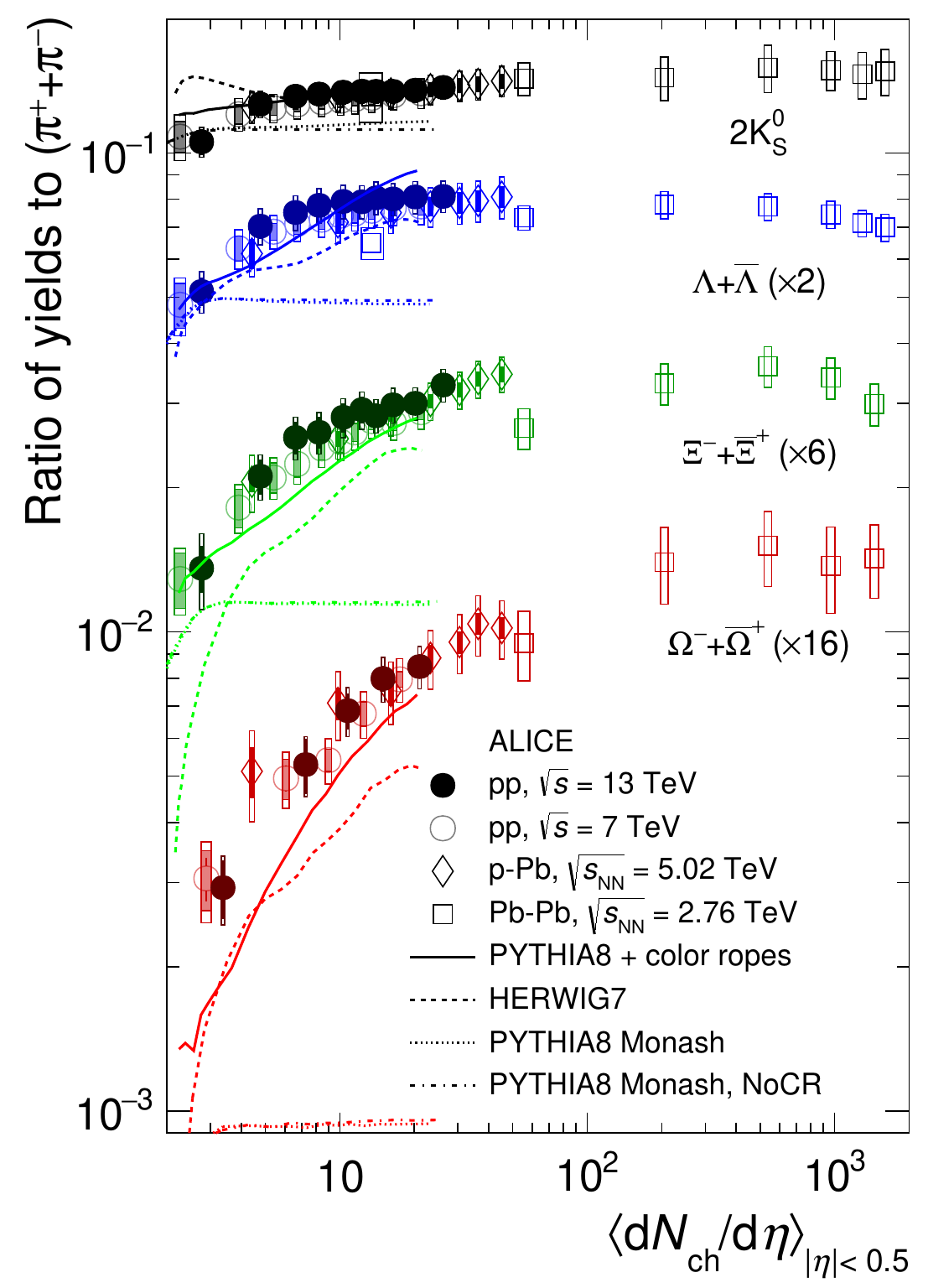}
    \caption{Integrated strange hadron-to-pion ratios as a function of \mdNde measured in pp, p--Pb, and Pb--Pb collisions. The open (shaded) boxes around markers represent full (multiplicity uncorrelated) systematic uncertainties. Different lines represent predictions from different MC generators for pp collisions at \cme{13}. References from~\cite{Adam:2015vsf,ALICE:2017jyt,Acharya:2018orn,Acharya:2019kyh}.}
    \label{fig:StrangeVsMulti}
  \end{center}
\end{figure}

The statistical uncertainties of \dNdy and \mpT are calculated by coherently shifting the central values of each spectra point by a fraction of its statistical uncertainty. The fraction is randomly drawn from Gaussian distribution and new values of integrated yields and mean transverse momenta are calculated. The procedure is repeated 1000 times to calculate the standard deviations of \dNdy and \mpT, which are then used as the statistical uncertainties.
To estimate the systematic uncertainty on the integrated yields, the spectra points are moved to maximal/minimal values allowed by their respective systematic uncertainties before repeating the fit procedure. For \mpT, each point of the spectra is shifted to the upper/lower edge of the corresponding \pT bin to obtain the hardest/softest particle distribution. The largest differences to the nominal yield and \mpT values are combined with the extrapolation uncertainties to calculate the total systematic uncertainties.  The kaon- and proton-to-pion integrated yield ratios measured in pp collisions at \cme{13} are found to be in a good agreement within systematic uncertainties with those measured in pp, p--Pb, and Pb--Pb collisions at \cmenn{\textrm{7, 5.02, and 2.76}}, respectively, as shown in \fig{IntPartRatioVsMulti}. In addition, with the availability of (multi)strange hadron yields~\cite{Acharya:2019kyh} we can study the relative abundances of hyperons to pions, and the results are shown in~\fig{StrangeVsMulti}.
We find that the (multi)strange hadron-to-pion ratios measured in pp collisions at \cme{13} are in good agreement to those measured at \cme{7} and similar \mdNde.
This indicates that hadrochemistry at LHC energies scales with charged-particle multiplicity density in a uniform way, despite the colliding system or collision energy.

The description of hadron-to-pion ratio factorization with multiplicity at lower center-of-mass energies in MC generators has been previously shown to be qualitative at best~\cite{Acharya:2018orn}. In fact, both \pythiaeight with color reconnection and HERWIG~7~\cite{Bellm:2015jjp,Bahr:2008pv} predict no evolution of the ratios with \mdNde. In this paper, we consider more recent versions of the two MC generators. In particular, the hadronization in \pythiaeight now considers overlapping color strings, which form color ropes with a larger effective string tension and are then allowed to interact with each other~\cite{Bierlich:2018xfw}. On the other hand, hadronization in HERWIG~7 now includes baryonic ropes -- a reconnection scheme that enhances the probability of partons forming a baryon~\cite{Bellm:2015jjp}. We find that both \pythiaeight and HERWIG~7 predict the enhancement of strange baryons which is more pronounced for hadrons with a larger strangeness content as shown in~\fig{StrangeVsMulti}. The largest quantitative differences are seen for $\Omega$/\pion ratio at the lowest multiplicity in pp collisions. The $\Xi$/\pion ratios are in a better agreement with \pythiaeight with color ropes, while HERWIG~7 shows a large deviation from the data at low \mdNde. Finally, $\Lambda$/\pion ratios are well described by HERWIG~7, while \pythiaeight with color ropes predicts an increasing trend in the whole multiplicity range available and overestimates the ratio at the highest multiplicities. Overall, the agreement between MC generators and measured hadron-to-pion ratios become worse for particles with a larger strangeness content.  This might point to the need of a further refinement of MC generator tuning, as similar trends are already observed for e$^{+}$e$^{-}$ data~\cite{Bierlich:2015rha}.

The integrated \kaon/\pion yield ratio shown in \fig{IntPartRatioVsMulti} at high multiplicity pp collisions are captured by \pythiaeight ropes and HERWIG~7, but the latter predicts a peak-like structure at low \mdNde which is not observed in the data.
The predictions from \pythiaeight Monash tune are inconsistent with the measured \kaon/\pion ratios in pp collisions at \cme{13}, whether color reconnection is considered or not. The quantitative description of \pr/\pion ratio is given only by HERWIG~7, while all considered versions of \pythiaeight overpredict the data. Moreover, \pythiaeight with color ropes predicts an increase of the \pr/\pion ratio with \mdNde, which could be attributed to the enhanced production of strange- and di-quark in the rope fragmentation. Overall, we conclude that none of the models considered provide a consistent description of the data.
\begin{figure}[b]
  \begin{center}
    \includegraphics[width=.95\textwidth]{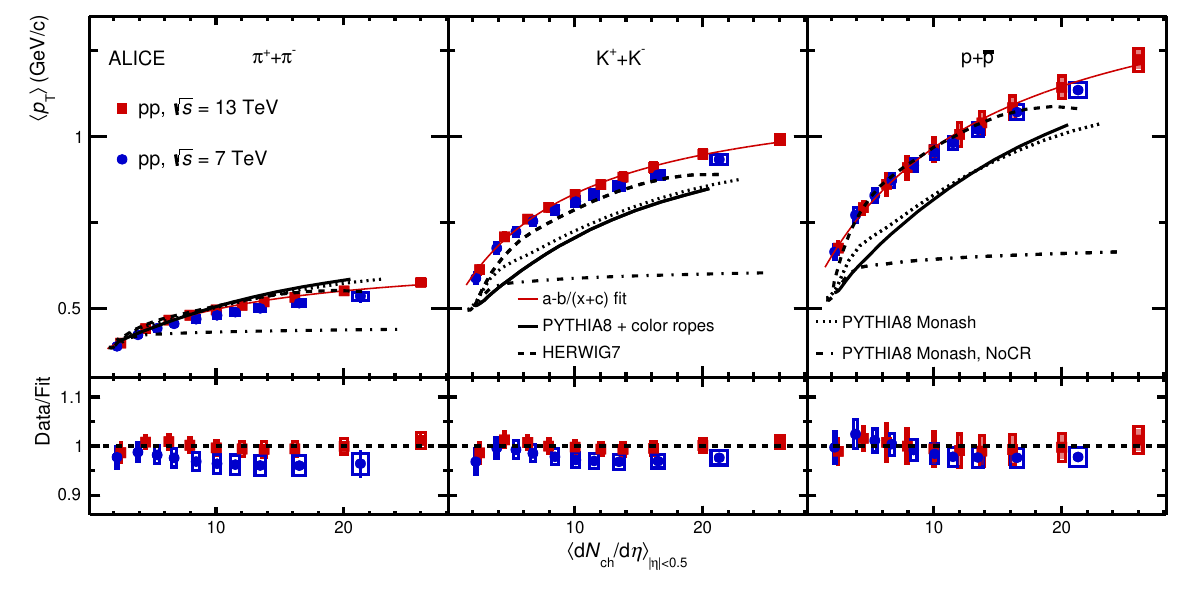}
    \caption{Upper panels: average transverse momenta of \pion, \kaon, and \pr as a function of charged-particle multiplicity density measured in pp collisions at \cme{\textrm{7 and 13}}. The red solid line represents $a-b(c-\mdNde)^{-1}$ fit to the 13~TeV data to guide the eye. Open (shaded) boxes represent total (multiplicity uncorrelated) systematic uncertainties. Black lines represent predictions from different MC generators for pp collisions at \cme{13}. Bottom panels: ratios of \mpT to the fits. Data at \cme{7} are from~\cite{Acharya:2018orn}.}
    \label{fig:MeanPt}
  \end{center}
\end{figure}

The average transverse momenta of identified particles are found to increase with multiplicity in pp collisions at \cme{\textrm{7 and 13}} as shown in \fig{MeanPt}. A clear mass ordering is observed among the particle species considered, where protons have the largest \mpT. Similar observations have been previously reported in pp~\cite{Chatrchyan:2012qb} and p--Pb~\cite{Abelev:2013haa} collisions at lower energies and for strange hadrons in pp collisions at \cme{13}~\cite{Acharya:2019kyh}. The solid red line in \fig{MeanPt} represents a fit of the form $a - b/(c-\mdNde)$ to the \cme{13} data, which is then used for a better comparison of \mpT between the two center-of-mass energies, see lower panels of the same figure. We find a small hint of an increase with \cmes for similar multiplicities for \pion, while the \mpT of protons is similar at the two center-of-mass energies. Note that similar observations have been already reported in~\cite{Acharya:2019kyh}, where spectra of K$^{0}_{\rm s}$ were found to become harder with \cmes at similar multiplicities. In addition, we find that \pythiaeight Monash tune with color reconnection, HERWIG~7, and \pythiaeight with ropes give a very good description of \pion \mpT evolution with \mdNde. This is expected as pions are the most abundant particles produced in collisions, and the three generators are tuned to explicitly to describe the \mpT of charged-particles. On the other hand, we observe that the \mpT of \kaon and \pr are well described only by HERWIG~7, while \pythiaeight with rope implementation underestimates the \mpT in the whole \mdNde range considered. This could be understood considering that the additional energy available during the rope fragmentation predominantly enhances the production of heavier hadrons at low \pT.

%% file: Inputs/Summary.tex
We have studied \pion, \kaon, and \pr production as a function of multiplicity in pp collisions at \cme{13}. To avoid auto-correlation biases, the event classification has been based on multiplicity measurements at forward (backward) pseudorapidity, while event activity \mdNde has been correspondingly estimated at central pseudorapidities, $|\eta|<0.5$. We find that hadron \pT spectra become harder with multiplicity, and the effect is more pronounced for heavier particles. The hardening of the spectra is predicted by \pythiaeight with rope hadronization, \pythiaeight Monash with color reconnection, and HERWIG7 MC generators. In addition, all three generators provide a quantitative description of \pion \mpT, while \kaon and \pr are described qualitatively only by HERWIG7. At high \pT $(\gtrsim 8\,\GeVc)$ we find that spectral shapes become independent of \mdNde as predicted by pQCD calculations~\cite{Kretzer:2000yf}.

The measured \pT-differential \kaon/\pion ratios show no evolution with multiplicity in the \pT range considered. In contrast to this, a depletion (enhancement, saturation) is visible for the \pr/\pion ratios at low (intermediate, high) \pT. In addition, we find that the ratios measured in pp collisions at \cme{13} are consistent with those measured at \cme{7}. The saturation at high \pT is captured by \pythiaeight Monash tunes, while HERWIG~7 and \pythiaeight with color ropes show signs of enhancement. While some of the most common MC generators capture the trends seen in the \pT-differential \kaon/\pion and \pr/\pion ratios, it is interesting to see that none of them provides a consistent description of the data and predict the absolute values of the ratios at high \pt.

The study of hadron \pT spectra in the context of the Blast-Wave model reveals that the kinetic freeze-out temperature \Tkin, average expansion velocity \mbeta, and the velocity profile exponent $n$ show little or no dependence on the center-of-mass energy and are consistent within uncertainties with those extracted from particle spectra measured in pp collisions at \cme{7}~\cite{Acharya:2018orn}. On the other hand, we observe a strong dependence of the extracted parameters on \mdNde.

The \pT-integrated hadron-to-pion ratios as a function of multiplicity show no center-of-mass dependence and the measurement in pp collisions at \cme{13} are compatible to those in pp, p--Pb, and Pb--Pb collisions at \cmenn{\textrm{7, 5.02, and 2.76}}, respectively.
This suggests that, at the LHC energies, the chemical composition of primary hadrons scales with charged-particle multiplicity density in a uniform way, despite the colliding system and collision energy.
Comparisons of the integrated hadron-to-pion ratios to the predictions from MC generators show that \pythiaeight with color ropes provides the best description of (multi)strange hadrons, but overestimates the measured \pr/\pion ratio. HERWIG7 also captures the evolution of the ratios with \mdNde, but underestimates the absolute values of $\Xi$/\pion and $\Omega$/\pion. Overall, none of the generators are able to provide a consistent quantitative description of the measured hadron-to-pion ratios.

%% file: fa_2020-02-05.tex

The ALICE Collaboration would like to thank all its engineers and technicians for their invaluable contributions to the construction of the experiment and the CERN accelerator teams for the outstanding performance of the LHC complex.
The ALICE Collaboration gratefully acknowledges the resources and support provided by all Grid centres and the Worldwide LHC Computing Grid (WLCG) collaboration.
The ALICE Collaboration acknowledges the following funding agencies for their support in building and running the ALICE detector:
A. I. Alikhanyan National Science Laboratory (Yerevan Physics Institute) Foundation (ANSL), State Committee of Science and World Federation of Scientists (WFS), Armenia;
Austrian Academy of Sciences, Austrian Science Fund (FWF): [M 2467-N36] and Nationalstiftung f\"{u}r Forschung, Technologie und Entwicklung, Austria;
Ministry of Communications and High Technologies, National Nuclear Research Center, Azerbaijan;
Conselho Nacional de Desenvolvimento Cient\'{\i}fico e Tecnol\'{o}gico (CNPq), Financiadora de Estudos e Projetos (Finep), Funda\c{c}\~{a}o de Amparo \`{a} Pesquisa do Estado de S\~{a}o Paulo (FAPESP) and Universidade Federal do Rio Grande do Sul (UFRGS), Brazil;
Ministry of Education of China (MOEC) , Ministry of Science \& Technology of China (MSTC) and National Natural Science Foundation of China (NSFC), China;
Ministry of Science and Education and Croatian Science Foundation, Croatia;
Centro de Aplicaciones Tecnol\'{o}gicas y Desarrollo Nuclear (CEADEN), Cubaenerg\'{\i}a, Cuba;
Ministry of Education, Youth and Sports of the Czech Republic, Czech Republic;
The Danish Council for Independent Research | Natural Sciences, the VILLUM FONDEN and Danish National Research Foundation (DNRF), Denmark;
Helsinki Institute of Physics (HIP), Finland;
Commissariat \`{a} l'Energie Atomique (CEA), Institut National de Physique Nucl\'{e}aire et de Physique des Particules (IN2P3) and Centre National de la Recherche Scientifique (CNRS) and R\'{e}gion des  Pays de la Loire, France;
Bundesministerium f\"{u}r Bildung und Forschung (BMBF) and GSI Helmholtzzentrum f\"{u}r Schwerionenforschung GmbH, Germany;
General Secretariat for Research and Technology, Ministry of Education, Research and Religions, Greece;
National Research, Development and Innovation Office, Hungary;
Department of Atomic Energy Government of India (DAE), Department of Science and Technology, Government of India (DST), University Grants Commission, Government of India (UGC) and Council of Scientific and Industrial Research (CSIR), India;
Indonesian Institute of Science, Indonesia;
Centro Fermi - Museo Storico della Fisica e Centro Studi e Ricerche Enrico Fermi and Istituto Nazionale di Fisica Nucleare (INFN), Italy;
Institute for Innovative Science and Technology , Nagasaki Institute of Applied Science (IIST), Japanese Ministry of Education, Culture, Sports, Science and Technology (MEXT) and Japan Society for the Promotion of Science (JSPS) KAKENHI, Japan;
Consejo Nacional de Ciencia (CONACYT) y Tecnolog\'{i}a, through Fondo de Cooperaci\'{o}n Internacional en Ciencia y Tecnolog\'{i}a (FONCICYT) and Direcci\'{o}n General de Asuntos del Personal Academico (DGAPA), Mexico;
Nederlandse Organisatie voor Wetenschappelijk Onderzoek (NWO), Netherlands;
The Research Council of Norway, Norway;
Commission on Science and Technology for Sustainable Development in the South (COMSATS), Pakistan;
Pontificia Universidad Cat\'{o}lica del Per\'{u}, Peru;
Ministry of Science and Higher Education and National Science Centre, Poland;
Korea Institute of Science and Technology Information and National Research Foundation of Korea (NRF), Republic of Korea;
Ministry of Education and Scientific Research, Institute of Atomic Physics and Ministry of Research and Innovation and Institute of Atomic Physics, Romania;
Joint Institute for Nuclear Research (JINR), Ministry of Education and Science of the Russian Federation, National Research Centre Kurchatov Institute, Russian Science Foundation and Russian Foundation for Basic Research, Russia;
Ministry of Education, Science, Research and Sport of the Slovak Republic, Slovakia;
National Research Foundation of South Africa, South Africa;
Swedish Research Council (VR) and Knut \& Alice Wallenberg Foundation (KAW), Sweden;
European Organization for Nuclear Research, Switzerland;
Suranaree University of Technology (SUT), National Science and Technology Development Agency (NSDTA) and Office of the Higher Education Commission under NRU project of Thailand, Thailand;
Turkish Atomic Energy Agency (TAEK), Turkey;
National Academy of  Sciences of Ukraine, Ukraine;
Science and Technology Facilities Council (STFC), United Kingdom;
National Science Foundation of the United States of America (NSF) and United States Department of Energy, Office of Nuclear Physics (DOE NP), United States of America.

%% file: 2020-02-05-Alice_Authorlist_2020-02-05.tex

\begingroup
\small
\begin{flushleft}
S.~Acharya\Irefn{org141}\And 
D.~Adamov\'{a}\Irefn{org94}\And 
A.~Adler\Irefn{org74}\And 
J.~Adolfsson\Irefn{org80}\And 
M.M.~Aggarwal\Irefn{org99}\And 
G.~Aglieri Rinella\Irefn{org33}\And 
M.~Agnello\Irefn{org30}\And 
N.~Agrawal\Irefn{org10}\textsuperscript{,}\Irefn{org53}\And 
Z.~Ahammed\Irefn{org141}\And 
S.~Ahmad\Irefn{org16}
% ...
\And
\renewcommand\labelenumi{\textsuperscript{\theenumi}~}

\section*{Affiliation notes}
\renewcommand\theenumi{\roman{enumi}}
\begin{Authlist}
\item \Adef{org*}Deceased
\item \Adef{orgI}Italian National Agency for New Technologies, Energy and Sustainable Economic Development (ENEA), Bologna, Italy
\item \Adef{orgII}Dipartimento DET del Politecnico di Torino, Turin, Italy
\item \Adef{orgIII}M.V. Lomonosov Moscow State University, D.V. Skobeltsyn Institute of Nuclear, Physics, Moscow, Russia
\item \Adef{orgIV}Department of Applied Physics, Aligarh Muslim University, Aligarh, India
\item \Adef{orgV}Institute of Theoretical Physics, University of Wroclaw, Poland
\end{Authlist}

\section*{Collaboration Institutes}
\renewcommand\theenumi{\arabic{enumi}~}
\begin{Authlist}
\item \Idef{org1}A.I. Alikhanyan National Science Laboratory (Yerevan Physics Institute) Foundation, Yerevan, Armenia
\item \Idef{org2}Bogolyubov Institute for Theoretical Physics, National Academy of Sciences of Ukraine, Kiev, Ukraine
\item \Idef{org3}Bose Institute, Department of Physics  and Centre for Astroparticle Physics and Space Science (CAPSS), Kolkata, India
\item \Idef{org4}Budker Institute for Nuclear Physics, Novosibirsk, Russia
\item \Idef{org5}California Polytechnic State University, San Luis Obispo, California, United States
\item \Idef{org6}Central China Normal University, Wuhan, China
\item \Idef{org7}Centre de Calcul de l'IN2P3, Villeurbanne, Lyon, France
\item \Idef{org8}Centro de Aplicaciones Tecnol\'{o}gicas y Desarrollo Nuclear (CEADEN), Havana, Cuba
\item \Idef{org9}Centro de Investigaci\'{o}n y de Estudios Avanzados (CINVESTAV), Mexico City and M\'{e}rida, Mexico
\item \Idef{org10}Centro Fermi - Museo Storico della Fisica e Centro Studi e Ricerche ``Enrico Fermi', Rome, Italy
\item \Idef{org11}Chicago State University, Chicago, Illinois, United States
\item \Idef{org12}China Institute of Atomic Energy, Beijing, China
\item \Idef{org13}Comenius University Bratislava, Faculty of Mathematics, Physics and Informatics, Bratislava, Slovakia
\item \Idef{org14}COMSATS University Islamabad, Islamabad, Pakistan
\item \Idef{org15}Creighton University, Omaha, Nebraska, United States
\item \Idef{org16}Department of Physics, Aligarh Muslim University, Aligarh, India
\item \Idef{org17}Department of Physics, Pusan National University, Pusan, Republic of Korea
\item \Idef{org18}Department of Physics, Sejong University, Seoul, Republic of Korea
\item \Idef{org19}Department of Physics, University of California, Berkeley, California, United States
\item \Idef{org20}Department of Physics, University of Oslo, Oslo, Norway
\item \Idef{org21}Department of Physics and Technology, University of Bergen, Bergen, Norway
\item \Idef{org22}Dipartimento di Fisica dell'Universit\`{a} 'La Sapienza' and Sezione INFN, Rome, Italy
\item \Idef{org23}Dipartimento di Fisica dell'Universit\`{a} and Sezione INFN, Cagliari, Italy
\item \Idef{org24}Dipartimento di Fisica dell'Universit\`{a} and Sezione INFN, Trieste, Italy
\item \Idef{org25}Dipartimento di Fisica dell'Universit\`{a} and Sezione INFN, Turin, Italy
\item \Idef{org26}Dipartimento di Fisica e Astronomia dell'Universit\`{a} and Sezione INFN, Bologna, Italy
\item \Idef{org27}Dipartimento di Fisica e Astronomia dell'Universit\`{a} and Sezione INFN, Catania, Italy
\item \Idef{org28}Dipartimento di Fisica e Astronomia dell'Universit\`{a} and Sezione INFN, Padova, Italy
\item \Idef{org29}Dipartimento di Fisica `E.R.~Caianiello' dell'Universit\`{a} and Gruppo Collegato INFN, Salerno, Italy
\item \Idef{org30}Dipartimento DISAT del Politecnico and Sezione INFN, Turin, Italy
\item \Idef{org31}Dipartimento di Scienze e Innovazione Tecnologica dell'Universit\`{a} del Piemonte Orientale and INFN Sezione di Torino, Alessandria, Italy
\item \Idef{org32}Dipartimento Interateneo di Fisica `M.~Merlin' and Sezione INFN, Bari, Italy
\item \Idef{org33}European Organization for Nuclear Research (CERN), Geneva, Switzerland
\item \Idef{org34}Faculty of Electrical Engineering, Mechanical Engineering and Naval Architecture, University of Split, Split, Croatia
\item \Idef{org35}Faculty of Engineering and Science, Western Norway University of Applied Sciences, Bergen, Norway
\item \Idef{org36}Faculty of Nuclear Sciences and Physical Engineering, Czech Technical University in Prague, Prague, Czech Republic
\item \Idef{org37}Faculty of Science, P.J.~\v{S}af\'{a}rik University, Ko\v{s}ice, Slovakia
\item \Idef{org38}Frankfurt Institute for Advanced Studies, Johann Wolfgang Goethe-Universit\"{a}t Frankfurt, Frankfurt, Germany
\item \Idef{org39}Fudan University, Shanghai, China
\item \Idef{org40}Gangneung-Wonju National University, Gangneung, Republic of Korea
\item \Idef{org41}Gauhati University, Department of Physics, Guwahati, India
\item \Idef{org42}Helmholtz-Institut f\"{u}r Strahlen- und Kernphysik, Rheinische Friedrich-Wilhelms-Universit\"{a}t Bonn, Bonn, Germany
\item \Idef{org43}Helsinki Institute of Physics (HIP), Helsinki, Finland
\item \Idef{org44}High Energy Physics Group,  Universidad Aut\'{o}noma de Puebla, Puebla, Mexico
\item \Idef{org45}Hiroshima University, Hiroshima, Japan
\item \Idef{org46}Hochschule Worms, Zentrum  f\"{u}r Technologietransfer und Telekommunikation (ZTT), Worms, Germany
\item \Idef{org47}Horia Hulubei National Institute of Physics and Nuclear Engineering, Bucharest, Romania
\item \Idef{org48}Indian Institute of Technology Bombay (IIT), Mumbai, India
\item \Idef{org49}Indian Institute of Technology Indore, Indore, India
\item \Idef{org50}Indonesian Institute of Sciences, Jakarta, Indonesia
\item \Idef{org51}INFN, Laboratori Nazionali di Frascati, Frascati, Italy
\item \Idef{org52}INFN, Sezione di Bari, Bari, Italy
\item \Idef{org53}INFN, Sezione di Bologna, Bologna, Italy
\item \Idef{org54}INFN, Sezione di Cagliari, Cagliari, Italy
\item \Idef{org55}INFN, Sezione di Catania, Catania, Italy
\item \Idef{org56}INFN, Sezione di Padova, Padova, Italy
\item \Idef{org57}INFN, Sezione di Roma, Rome, Italy
\item \Idef{org58}INFN, Sezione di Torino, Turin, Italy
\item \Idef{org59}INFN, Sezione di Trieste, Trieste, Italy
\item \Idef{org60}Inha University, Incheon, Republic of Korea
\item \Idef{org61}Institut de Physique Nucl\'{e}aire d'Orsay (IPNO), Institut National de Physique Nucl\'{e}aire et de Physique des Particules (IN2P3/CNRS), Universit\'{e} de Paris-Sud, Universit\'{e} Paris-Saclay, Orsay, France
\item \Idef{org62}Institute for Nuclear Research, Academy of Sciences, Moscow, Russia
\item \Idef{org63}Institute for Subatomic Physics, Utrecht University/Nikhef, Utrecht, Netherlands
\item \Idef{org64}Institute of Experimental Physics, Slovak Academy of Sciences, Ko\v{s}ice, Slovakia
\item \Idef{org65}Institute of Physics, Homi Bhabha National Institute, Bhubaneswar, India
\item \Idef{org66}Institute of Physics of the Czech Academy of Sciences, Prague, Czech Republic
\item \Idef{org67}Institute of Space Science (ISS), Bucharest, Romania
\item \Idef{org68}Institut f\"{u}r Kernphysik, Johann Wolfgang Goethe-Universit\"{a}t Frankfurt, Frankfurt, Germany
\item \Idef{org69}Instituto de Ciencias Nucleares, Universidad Nacional Aut\'{o}noma de M\'{e}xico, Mexico City, Mexico
\item \Idef{org70}Instituto de F\'{i}sica, Universidade Federal do Rio Grande do Sul (UFRGS), Porto Alegre, Brazil
\item \Idef{org71}Instituto de F\'{\i}sica, Universidad Nacional Aut\'{o}noma de M\'{e}xico, Mexico City, Mexico
\item \Idef{org72}iThemba LABS, National Research Foundation, Somerset West, South Africa
\item \Idef{org73}Jeonbuk National University, Jeonju, Republic of Korea
\item \Idef{org74}Johann-Wolfgang-Goethe Universit\"{a}t Frankfurt Institut f\"{u}r Informatik, Fachbereich Informatik und Mathematik, Frankfurt, Germany
\item \Idef{org75}Joint Institute for Nuclear Research (JINR), Dubna, Russia
\item \Idef{org76}Korea Institute of Science and Technology Information, Daejeon, Republic of Korea
\item \Idef{org77}KTO Karatay University, Konya, Turkey
\item \Idef{org78}Laboratoire de Physique Subatomique et de Cosmologie, Universit\'{e} Grenoble-Alpes, CNRS-IN2P3, Grenoble, France
\item \Idef{org79}Lawrence Berkeley National Laboratory, Berkeley, California, United States
\item \Idef{org80}Lund University Department of Physics, Division of Particle Physics, Lund, Sweden
\item \Idef{org81}Nagasaki Institute of Applied Science, Nagasaki, Japan
\item \Idef{org82}Nara Women{'}s University (NWU), Nara, Japan
\item \Idef{org83}National and Kapodistrian University of Athens, School of Science, Department of Physics , Athens, Greece
\item \Idef{org84}National Centre for Nuclear Research, Warsaw, Poland
\item \Idef{org85}National Institute of Science Education and Research, Homi Bhabha National Institute, Jatni, India
\item \Idef{org86}National Nuclear Research Center, Baku, Azerbaijan
\item \Idef{org87}National Research Centre Kurchatov Institute, Moscow, Russia
\item \Idef{org88}Niels Bohr Institute, University of Copenhagen, Copenhagen, Denmark
\item \Idef{org89}Nikhef, National institute for subatomic physics, Amsterdam, Netherlands
\item \Idef{org90}NRC Kurchatov Institute IHEP, Protvino, Russia
\item \Idef{org91}NRC \guillemotleft Kurchatov\guillemotright~Institute - ITEP, Moscow, Russia
\item \Idef{org92}NRNU Moscow Engineering Physics Institute, Moscow, Russia
\item \Idef{org93}Nuclear Physics Group, STFC Daresbury Laboratory, Daresbury, United Kingdom
\item \Idef{org94}Nuclear Physics Institute of the Czech Academy of Sciences, \v{R}e\v{z} u Prahy, Czech Republic
\item \Idef{org95}Oak Ridge National Laboratory, Oak Ridge, Tennessee, United States
\item \Idef{org96}Ohio State University, Columbus, Ohio, United States
\item \Idef{org97}Petersburg Nuclear Physics Institute, Gatchina, Russia
\item \Idef{org98}Physics department, Faculty of science, University of Zagreb, Zagreb, Croatia
\item \Idef{org99}Physics Department, Panjab University, Chandigarh, India
\item \Idef{org100}Physics Department, University of Jammu, Jammu, India
\item \Idef{org101}Physics Department, University of Rajasthan, Jaipur, India
\item \Idef{org102}Physikalisches Institut, Eberhard-Karls-Universit\"{a}t T\"{u}bingen, T\"{u}bingen, Germany
\item \Idef{org103}Physikalisches Institut, Ruprecht-Karls-Universit\"{a}t Heidelberg, Heidelberg, Germany
\item \Idef{org104}Physik Department, Technische Universit\"{a}t M\"{u}nchen, Munich, Germany
\item \Idef{org105}Politecnico di Bari, Bari, Italy
\item \Idef{org106}Research Division and ExtreMe Matter Institute EMMI, GSI Helmholtzzentrum f\"ur Schwerionenforschung GmbH, Darmstadt, Germany
\item \Idef{org107}Rudjer Bo\v{s}kovi\'{c} Institute, Zagreb, Croatia
\item \Idef{org108}Russian Federal Nuclear Center (VNIIEF), Sarov, Russia
\item \Idef{org109}Saha Institute of Nuclear Physics, Homi Bhabha National Institute, Kolkata, India
\item \Idef{org110}School of Physics and Astronomy, University of Birmingham, Birmingham, United Kingdom
\item \Idef{org111}Secci\'{o}n F\'{\i}sica, Departamento de Ciencias, Pontificia Universidad Cat\'{o}lica del Per\'{u}, Lima, Peru
\item \Idef{org112}St. Petersburg State University, St. Petersburg, Russia
\item \Idef{org113}Stefan Meyer Institut f\"{u}r Subatomare Physik (SMI), Vienna, Austria
\item \Idef{org114}SUBATECH, IMT Atlantique, Universit\'{e} de Nantes, CNRS-IN2P3, Nantes, France
\item \Idef{org115}Suranaree University of Technology, Nakhon Ratchasima, Thailand
\item \Idef{org116}Technical University of Ko\v{s}ice, Ko\v{s}ice, Slovakia
\item \Idef{org117}Technische Universit\"{a}t M\"{u}nchen, Excellence Cluster 'Universe', Munich, Germany
\item \Idef{org118}The Henryk Niewodniczanski Institute of Nuclear Physics, Polish Academy of Sciences, Cracow, Poland
\item \Idef{org119}The University of Texas at Austin, Austin, Texas, United States
\item \Idef{org120}Universidad Aut\'{o}noma de Sinaloa, Culiac\'{a}n, Mexico
\item \Idef{org121}Universidade de S\~{a}o Paulo (USP), S\~{a}o Paulo, Brazil
\item \Idef{org122}Universidade Estadual de Campinas (UNICAMP), Campinas, Brazil
\item \Idef{org123}Universidade Federal do ABC, Santo Andre, Brazil
\item \Idef{org124}University of Cape Town, Cape Town, South Africa
\item \Idef{org125}University of Houston, Houston, Texas, United States
\item \Idef{org126}University of Jyv\"{a}skyl\"{a}, Jyv\"{a}skyl\"{a}, Finland
\item \Idef{org127}University of Liverpool, Liverpool, United Kingdom
\item \Idef{org128}University of Science and Technology of China, Hefei, China
\item \Idef{org129}University of South-Eastern Norway, Tonsberg, Norway
\item \Idef{org130}University of Tennessee, Knoxville, Tennessee, United States
\item \Idef{org131}University of the Witwatersrand, Johannesburg, South Africa
\item \Idef{org132}University of Tokyo, Tokyo, Japan
\item \Idef{org133}University of Tsukuba, Tsukuba, Japan
\item \Idef{org134}Universit\'{e} Clermont Auvergne, CNRS/IN2P3, LPC, Clermont-Ferrand, France
\item \Idef{org135}Universit\'{e} de Lyon, Universit\'{e} Lyon 1, CNRS/IN2P3, IPN-Lyon, Villeurbanne, Lyon, France
\item \Idef{org136}Universit\'{e} de Strasbourg, CNRS, IPHC UMR 7178, F-67000 Strasbourg, France, Strasbourg, France
\item \Idef{org137}Universit\'{e} Paris-Saclay Centre d'Etudes de Saclay (CEA), IRFU, D\'{e}partment de Physique Nucl\'{e}aire (DPhN), Saclay, France
\item \Idef{org138}Universit\`{a} degli Studi di Foggia, Foggia, Italy
\item \Idef{org139}Universit\`{a} degli Studi di Pavia, Pavia, Italy
\item \Idef{org140}Universit\`{a} di Brescia, Brescia, Italy
\item \Idef{org141}Variable Energy Cyclotron Centre, Homi Bhabha National Institute, Kolkata, India
\item \Idef{org142}Warsaw University of Technology, Warsaw, Poland
\item \Idef{org143}Wayne State University, Detroit, Michigan, United States
\item \Idef{org144}Westf\"{a}lische Wilhelms-Universit\"{a}t M\"{u}nster, Institut f\"{u}r Kernphysik, M\"{u}nster, Germany
\item \Idef{org145}Wigner Research Centre for Physics, Budapest, Hungary
\item \Idef{org146}Yale University, New Haven, Connecticut, United States
\item \Idef{org147}Yonsei University, Seoul, Republic of Korea
\end{Authlist}
\endgroup